\documentclass[final,5p,times,twocolumn]{elsarticle}

\usepackage{amssymb}
\usepackage{amsmath}

\usepackage{natbib}
\biboptions{numbers,sort&compress}
\usepackage[colorlinks=true,linkcolor=blue,citecolor=blue,urlcolor=blue]{hyperref}
\usepackage{subfigure}
\usepackage{graphicx}
\usepackage{latexsym}
\usepackage{bm} 
\usepackage{amsfonts}
\usepackage{soul,color,xcolor} 
\usepackage{stfloats} 
\usepackage{booktabs} 
\usepackage{tabularx}
\usepackage{tabu}
\usepackage{multirow}
\usepackage{url}
\usepackage{geometry}

\usepackage{threeparttable} 
\usepackage{makecell} 
\usepackage{bbding} 
\usepackage{algorithm}
\usepackage{algorithmic}

\journal{Elsevier}

\begin{document}
	
	\begin{frontmatter}
		
		
		
		
		\title{When are supercapacitors practically feasible in electric vehicles? \\ A multi-dimensional HESS techno-economic evaluation}
		
		
%
%
%

		\author[a]{Yue Wu\fnref{equal}}
		\author[a]{Ziqing Xia\fnref{equal}}
		\author[b]{Shaokun Li}
		\author[a]{Heng Li}
		\author[c]{Shengyu Tao}
		\author[b]{Zhiwu Huang\corref{cor1}}
		
		\fntext[equal]{These authors contributed equally.}
		\cortext[cor1]{Corresponding author.}\ead{hzw@csu.edu.cn}
		
		\address[a]{School of Electronic Information, Central South University,	Changsha 410075, China}
		\address[b]{School of Automation, Central South University,	Changsha 410083, China}
		\address[c]{Department of Electrical Engineering, Chalmers University of Technology, Gothenburg 41296, Sweden}
		
		
		\begin{abstract}
			While the hybrid energy storage system (HESS) can theoretically mitigate battery degradation in electric vehicles, its practical implementation remains highly limited. To delineate the specific scenarios and application boundaries where supercapacitors remain feasible, this study proposes a multi-dimensional techno-economic feasibility evaluation framework. First, a cross-vehicle sizing method based on dynamic programming is established to quantify physical mass-volume packaging constraints and identify feasible supercapacitor candidates across different vehicle types. Building upon the optimal sizing parameters derived from the battery aging Pareto front, an expert-guided deep reinforcement learning energy management strategy is integrated to yield near-optimal online performance, ensuring a fair life-cycle economic assessment. Finally, a comprehensive feasibility matrix is constructed to systematically evaluate mass, volume, battery lifespan, additional supercapacitor costs, operation cost, future energy storage prices, and the influence of emerging solid-state batteries. Results reveal that city buses remain the most promising vehicle type for HESS due to minimal additional costs and sufficient packaging space. Current mass-volume penalties and limited economic benefits hinder HESS application in passenger vehicles and heavy-duty trucks, respectively. This situation may only improve if supercapacitor prices drop significantly in the future. Quantified analysis shows that higher load-frequency characteristics generally favor HESS benefits within the same vehicle platform, while the overall techno-economic feasibility across vehicles is jointly influenced by load characteristics and vehicle configuration. Furthermore, looking toward the 2030+ solid-state battery era, we highlight that integrating increasingly affordable supercapacitors can provide substantial asset protection leverage.
		\end{abstract}
		
		
	\end{frontmatter}
	
	
	\section{Introduction}
	%
	%
	%
	%
	
	Battery degradation induced by high-frequency and high-amplitude dynamic loads remains a critical challenge in electric vehicles. A hybrid energy storage system (HESS), which couples a high-energy-density battery pack with a high-power-density supercapacitor (SC) pack, has therefore been widely recognized as an effective solution \cite{xiong2018towards}. By absorbing transient peak currents \cite{wu2020adaptive,ccorapsiz2024pv}, the SC can mitigate severe power stress on the battery, thereby extending battery life and improving vehicle economy \cite{wang2020review}.
	
	\begin{table*}[!t]
	\resizebox{\textwidth}{34mm}{
		\begin{tabular}{|lllll}
			\hline
			\multicolumn{2}{|l}
			{\textbf
				{Nomenclature}} & \hspace*{3.8pt} &   & \multicolumn{1}{l|}{} \\
			$T_{w}$	& total wheel torque of the electric vehicle	&  & $V_{sc,cell}$ / $C_{sc,cell}$ & \multicolumn{1}{l|}{voltage/nominal capacitance of the SC cell} \\
			$m$ / $m_{veh}$	& mass of the electric vehicle/vehicle without HESS &  & $E_{sc,cell}$ / $r_{sc,cell}$ & \multicolumn{1}{l|}{stored energy/internal resistance of the SC cell} \\
			$m_{load}$ / $m_{sc}$	& load mass (passengers or cargo)/supercapacitor pack mass	&  & $m_{sc,cell}$ / Vol$_{sc,cell}$ & \multicolumn{1}{l|}{SC cell mass/volume} \\
			$v$ / $a$	& vehicle speed/acceleration	&  & SoC$_{sc}$ / $V_{sc,max}$ / $V_{sc,min}$ & \multicolumn{1}{l|}{state of charge/maximum/minimum voltage of the SC pack} \\
			$\theta$	& road surface angle	&  & $C_{sc}$ / Vol$_{sc}$ & \multicolumn{1}{l|}{capacitance/volume of the SC pack} \\
			$C_{r}$ / $C_{d}$	& rolling resistance/aerodynamic drag coefficient	&  & $E_{sc}$ / $I_{sc}$ & \multicolumn{1}{l|}{stored energy/current of the SC pack} \\
			$\rho$ / $A_{f}$	& air density/vehicle frontal area	&  & $P_{sc}$ / $P_{sc,total}$ & \multicolumn{1}{l|}{actual power/total power of the SC pack} \\
			$r_{w}$	& wheel radius	&  & $N_{s,bat}$ / $N_{p,bat}$ & \multicolumn{1}{l|}{battery cell serial/parallel number} \\
			$g$	& gravitational acceleration	&  & $N_{s,sc}$ / $N_{p,sc}$ & \multicolumn{1}{l|}{SC cell serial/parallel number} \\
			$T_{m}$ / $n_{m}$	& motor torque/rotational speed	&  & $N_{s,sc,i}$ / $N_{p,sc,max}$ & \multicolumn{1}{l|}{$i$-th feasible SC cell serial candidate/maximum SC cell parallel number} \\
			$i_{0}$ / $i_{1}$	& reduction ratio of gearbox/AMT	&  & $N_{p,sc,max,mass}$ / $N_{p,sc,max,vol}$ & \multicolumn{1}{l|}{maximum SC cell parallel number restricted by mass/volume} \\
			$P_{m}$ / $P_{d}$	& motor/DC side power demand &  & $k_{vol,bat}$ / $k_{m,bat}$ & \multicolumn{1}{l|}{volume/mass coefficient for battery packaging} \\
			$\eta_{m}$ / $\eta_{dcac}$	& efficiency of the motor/inverter	&  & $k_{vol,sc}$ / $k_{m,sc}$ & \multicolumn{1}{l|}{volume/mass coefficient for SC packaging} \\
			$V_{bat,oc,cell}$ / $V_{bat,cell}$	& open circuit voltage/voltage range of the battery cell	&  & $k_{DCDC,max}$ / $k_{DCDC,min}$ & \multicolumn{1}{l|}{maximum/minimum voltage boost ratio of the DC/DC converter} \\
			$Q_{bat,cell}$ / $E_{bat,cell}$	& nominal capacity/stored energy of the battery cell	&  &  &  \multicolumn{1}{l|}{} \\
			$r_{bat,cell}$ / $m_{bat,cell}$ / Vol$_{bat,cell}$	& internal resistance/mass/volume of the battery cell	&  & \multicolumn{2}{l|}{\textbf{Abbreviations}} \\
			SoC$_{bat}$	& state of charge of the battery pack	&  &  & \multicolumn{1}{l|}{}  \\
			$Q_{bat}$	& capacity of the battery pack	&  & HESS & \multicolumn{1}{l|}{hybrid energy storage system} \\
			$V_{bat,oc}$ / $V_{bat,pack,max}$ / $V_{bat,pack,min}$	& nominal/maximum/minimum voltage of the battery pack	&  & EV & \multicolumn{1}{l|}{electric vehicle} \\
			Vol$_{bat}$ / $m_{bat}$	& volume/mass of the battery pack	&  & DP & \multicolumn{1}{l|}{dynamic programming} \\
			$P_{bat}$ / $I_{bat}$	& actual power/current of the battery pack	&  & MPC & \multicolumn{1}{l|}{model predictive control} \\
			$Q_{loss}$	& battery capacity loss	&  & DRL & \multicolumn{1}{l|}{deep reinforcement learning} \\
			$price_{bat}$	& battery price	&  & TD3 & \multicolumn{1}{l|}{Twin Delayed Deep Deterministic Policy Gradient} \\
			$V_{sc,cell,max}$ / $V_{sc,cell,min}$	& maximum/minimum voltage of the SC cell	&  & BC & \multicolumn{1}{l|}{behavioral cloning} \\ \hline
		\end{tabular}
	}
	\end{table*}
	
	Despite extensive studies demonstrating these theoretical benefits over the past decade \cite{zhang2021hybrid,ramesh2024review}, the commercial adoption of HESS in practical electric vehicles remains limited. Recent reviews attribute this gap to the low energy density and high cost of SCs, together with additional converter, control, packaging, and system-integration complexity \cite{li2026toward,parvez2026supercapacitor}. Consequently, practical HESS implementation depends on three coupled issues: (1) optimal SC sizing, (2) effective onboard energy management, and (3) overall techno-economic feasibility.
	
	In terms of sizing, existing studies have balanced electrical requirements, initial investment, system mass, and battery degradation. Dynamic programming (DP) has been adopted for globally optimal power allocation in electric buses, enabling different SC combinations to be evaluated along the HESS life-cost Pareto frontier \cite{song2015optimization}. Genetic-algorithm-based sizing concurrently minimizes system weight, manufacturing cost, and battery capacity degradation \cite{zhang2017multiobjective}, while a universal double-layer method considers diverse HESS topologies \cite{nguyen2024universal}. Joint sizing and energy-management optimization has also been used to balance SC cost and battery degradation \cite{wang2021energy}. The additional SC mass and its resulting traction-power increase were explicitly considered in \cite{huang2022sizing}, indicating that the cost Pareto frontier is not monotonic. In recent years, application-specific extensions have addressed extreme-temperature light commercial vehicles \cite{pang2024optimal} and electric racing cars with lateral-dynamic constraints \cite{guan2025hierarchical}.
	
	Beyond sizing, onboard energy management is essential for translating the theoretical HESS benefit into real-time battery protection. Recent reviews have categorized existing strategies into filter-, model predictive control (MPC)-, and reinforcement learning-based methods \cite{he2024deep,gunther2025representative}. Filter-based strategies allocate the low- and high-frequency power components to the battery and SC, respectively \cite{wu2020adaptive}, while adaptive and converter-level current-shaping approaches further regulate battery stress \cite{ccorapsiz2022double,ccorapsiz2023overview}. MPC has been used to balance SC voltage regulation and battery current smoothing \cite{nguyen2022standalone}, and vehicle-cloud \cite{tang2024optimal} or multi-horizon MPC \cite{wu2023spatial} can exploit preview information to improve power allocation and battery life performance \cite{wu2024integrated}. Reinforcement learning \cite{xiong2018reinforcement}, incentive deep reinforcement learning (DRL) \cite{li2023incentive}, imitation learning-assisted DRL \cite{liu2025imitation}, and online policy updating \cite{guan2025towards} all provide alternative data-driven solutions for nonlinear online control and battery life extension.
	
	\begin{table*}[!t]
		\caption{Comparison of representative battery--supercapacitor HESS studies in electric vehicles.}
		\label{tab1}
		\centering
		\scriptsize
		\setlength{\tabcolsep}{2.5pt}
		\renewcommand{\arraystretch}{1.18}
		\begin{threeparttable}
			\begin{tabularx}{\textwidth}{@{}
					>{\raggedright\arraybackslash}p{1.25cm}
					>{\raggedright\arraybackslash}p{2.05cm}
					>{\raggedright\arraybackslash}p{1.55cm}
					>{\raggedright\arraybackslash}p{2.85cm}
					>{\raggedright\arraybackslash}X
					>{\raggedright\arraybackslash}p{3.65cm}
					>{\raggedright\arraybackslash}p{1.55cm}@{}}
				\toprule
				Literature
				& Vehicle
				& \makecell[l]{Mass/Volume}
				& Sizing method
				& EMS method
				& Feasibility analysis
				& \makecell[l]{Future battery\\technology} \\
				\midrule
				
				Ref. \cite{parvez2026supercapacitor}
				& EVs (Review)
				& --/--
				& --
				& Review of HESS topologies and EMSs
				& Techno-economic, life-cycle, and industrial-adoption review
				& No (SC roadmap) \\
				
				\midrule
				
				Ref. \cite{song2015optimization}
				& Bus
				& No/No
				& DP-based SC sizing
				& DP-extracted rule-based EMS
				& Life-cycle cost analysis
				& No \\
				
				\midrule
				
				Ref. \cite{zhang2017multiobjective}
				& Passenger vehicle
				& Yes/No
				& Multi-objective genetic algorithm
				& Wavelet-transform
				& Energy storage cost and battery life assessment
				& No \\
				
				\midrule
				
				Ref. \cite{nguyen2024universal}
				& Passenger vehicle
				& No/Yes
				& Double-layer NSGA-II
				& Alternating PMP
				& Techno-economic Pareto and installation-space analysis
				& No \\
				
				\midrule
				
				Ref. \cite{wang2021energy}
				& Not mentioned
				& No/No
				& MOEA/D-based sizing
				& Adaptive MPC with GWO--SVM adaptation
				& SC cost and battery degradation assessment
				& No \\
				
				\midrule
				
				Ref. \cite{huang2022sizing}
				& Passenger vehicle
				& Yes/No
				& DP-based SC-number traversal
				& Offline DP
				& Operation cost analysis with added-mass effect
				& No \\
				
				\midrule
				
				Ref. \cite{pang2024optimal}
				& Light-duty Truck
				& No/No
				& DP
				& DP baseline and EKF-assisted MPC
				& Life-cycle cost analysis at extreme temperatures
				& No \\
				
				\midrule
				
				Ref. \cite{guan2025hierarchical}
				& Racing car
				& Yes/No
				& Hierarchical Bayesian optimization
				& Numerical optimization
				& Racing-time assessment across tracks
				& No \\
				
				\midrule
				
				Ref. \cite{liu2025imitation}
				& Passenger vehicle
				& --/--
				& --
				& Adversarial imitation reinforcement learning with DP experts
				& Operation cost analysis
				& No \\
				
				\midrule
				
				Ref. \cite{guan2025towards}
				& Passenger vehicle
				& --/--
				& --
				& JS-divergence-triggered online DRL updating
				& Operation cost analysis
				& No \\
				
				\midrule
				
				Ref. \cite{zhu2021optimal}
				& Passenger vehicle
				& No/No
				& DP-based Joint sizing--energy management optimization
				& DP
				& Life-cycle cost sensitivity analysis
				& No \\
				
				\midrule
				
				Ref. \cite{wu2022hierarchical}
				& Passenger vehicle
				& --/--
				& --
				& Hierarchical MPC
				& Operation cost analysis
				& No \\
				
				\midrule
%
%
%
%
				
				Ref. \cite{song2018battery}
				& Bus
				& No/No
				& DP-based SC sizing
				& DP-derived rule-based EMS
				& Life-cycle cost analysis with battery price and temperature sensitivity
				& No (price trend only) \\
				
				\midrule
				
				This work
				& Sedan, SUV, Bus, and Truck
				& Yes/Yes
				& DP-based SC sizing with feasible candidates
				& BC-initialized TD3
				& Cross-vehicle multi-dimensional techno-economic analysis
				& Solid-state batteries \\
				\bottomrule
			\end{tabularx}
			
			\begin{tablenotes}
				\footnotesize
				\item[] ``Mass/Volume'' indicates whether the additional HESS mass and packaging volume are explicitly incorporated into the sizing or vehicle load calculation. ``Price trend only'' denotes battery price projection without modeling an emerging battery chemistry.
			\end{tablenotes}
		\end{threeparttable}
	\end{table*}
	
	\begin{figure*}[!t]
		\centering
		\includegraphics[width=16cm]{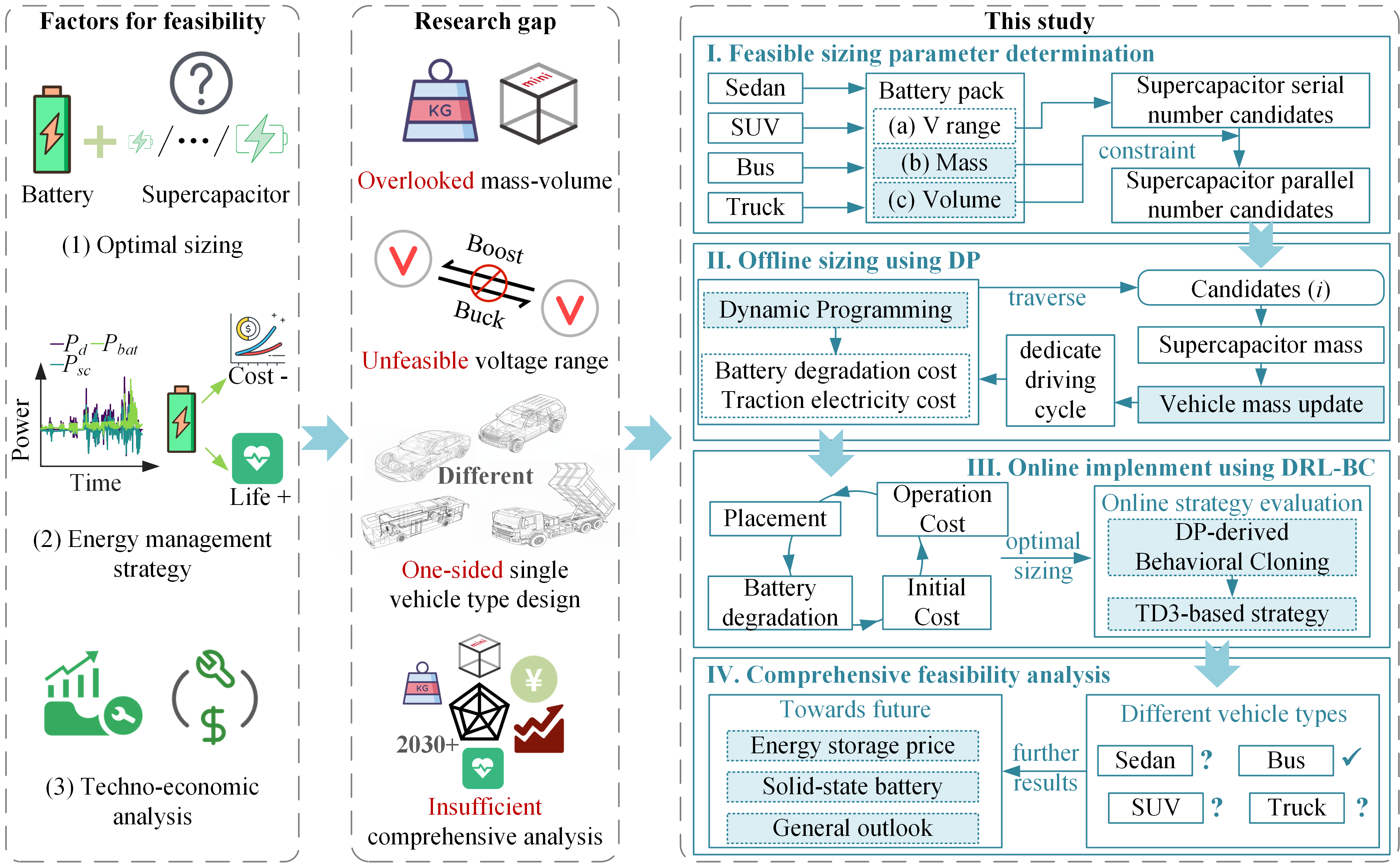}
		\caption{A multi-dimensional techno-economic evaluation for hybrid energy storage system in electric vehicles.} 
		\label{fig1_structure}
	\end{figure*}

	Building upon sizing and energy management, various studies have evaluated HESS feasibility through experimental implementation and life-cycle economic analysis. Comprehensive sensitivity analyses have quantified the effects of vehicle, system, and component parameters on HESS life-cycle cost \cite{zhu2021optimal}, while passenger-vehicle studies under car-following conditions have shown that battery degradation can dominate the operation cost \cite{wu2022hierarchical}. Physical implementation has been demonstrated using a full-scale test bench for urban commercial vehicles \cite{veneri2018experimental} and a 30kW semi-active experimental platform under standard driving cycles \cite{zhang2019experimental}. Meanwhile, declining battery prices can weaken the economic justification for adding SCs \cite{song2018battery}. Recent reviews have broadened this discussion to include topology selection, life-cycle impacts, industrial adoption, and future technology roadmaps \cite{parvez2026supercapacitor}.
	
	Despite these advances, it remains difficult to answer the question: \textbf{When are supercapacitors practically feasible in electric vehicles?} (1) Sizing, packaging, and control are often assessed separately, so feasible serial--parallel configurations, voltage compatibility, mass--volume penalties, and the resulting vehicle-load increase are not consistently propagated into online energy-management and economic evaluations; (2) Research based on a single vehicle type is often one-sided, failing to account for the diverse spatial constraints and load capacity margins across different vehicle types; (3) A unified cross-vehicle assessment that simultaneously quantifies battery life extension, operation cost, additional SC investment, mass, volume, future energy storage prices, and the influence of emerging solid-state batteries has not been reported.

	To address these gaps as listed in Table \ref{tab1}, a multi-dimensional HESS feasibility framework shown in Fig. \ref{fig1_structure} is established for four representative electric vehicles: Sedan, SUV, Bus, and Truck. Feasible SC serial--parallel candidates are first determined from the battery-voltage range and the mass--volume constraints; the SC mass is then included in the vehicle dynamics, and DP is used to optimize power allocation for every candidate. The optimal SC sizing is selected from the battery-aging Pareto front while accounting for placement and initial cost. Subsequently, an online energy management strategy is introduced to quantify the online performance, where Twin Delayed Deep Deterministic Policy Gradient (TD3) is initialized by behavioral cloning (BC). Afterward, battery degradation reduction, operation cost, energy storage price, mass, volume, and the future solid-state battery scenario are comprehensively assessed.
	
	The contributions of this study are listed below:
	
	(1) A cross-vehicle sizing method is established to explicitly quantify physical mass--volume packaging penalties and determine feasible SC configurations across different electric vehicle types.
	
	(2) An expert-guided DRL-based energy management strategy is introduced and compared with existing methods to ensure a fair online life-cycle economic feasibility assessment for different vehicle types.
	
	(3) A multi-dimensional techno-economic framework is constructed to reveal how vehicle constraints, load-frequency characteristics, energy storage prices, and future battery technologies jointly determine the practical application boundaries of HESS.

	This study is organized as follows. Section \ref{model} presents the HESS models and vehicle parameters. Section \ref{method} introduces the offline DP sizing method and the online DRL-based energy management strategies. Section \ref{result1} provides the comparative results, comprehensive feasibility analysis, application recommendations, and discussions. The conclusions are presented in Section \ref{conclusion}.
	
	\section{Preliminaries \label{model}}

\begin{figure}[!h]
		\centering
		\includegraphics[width=8cm]{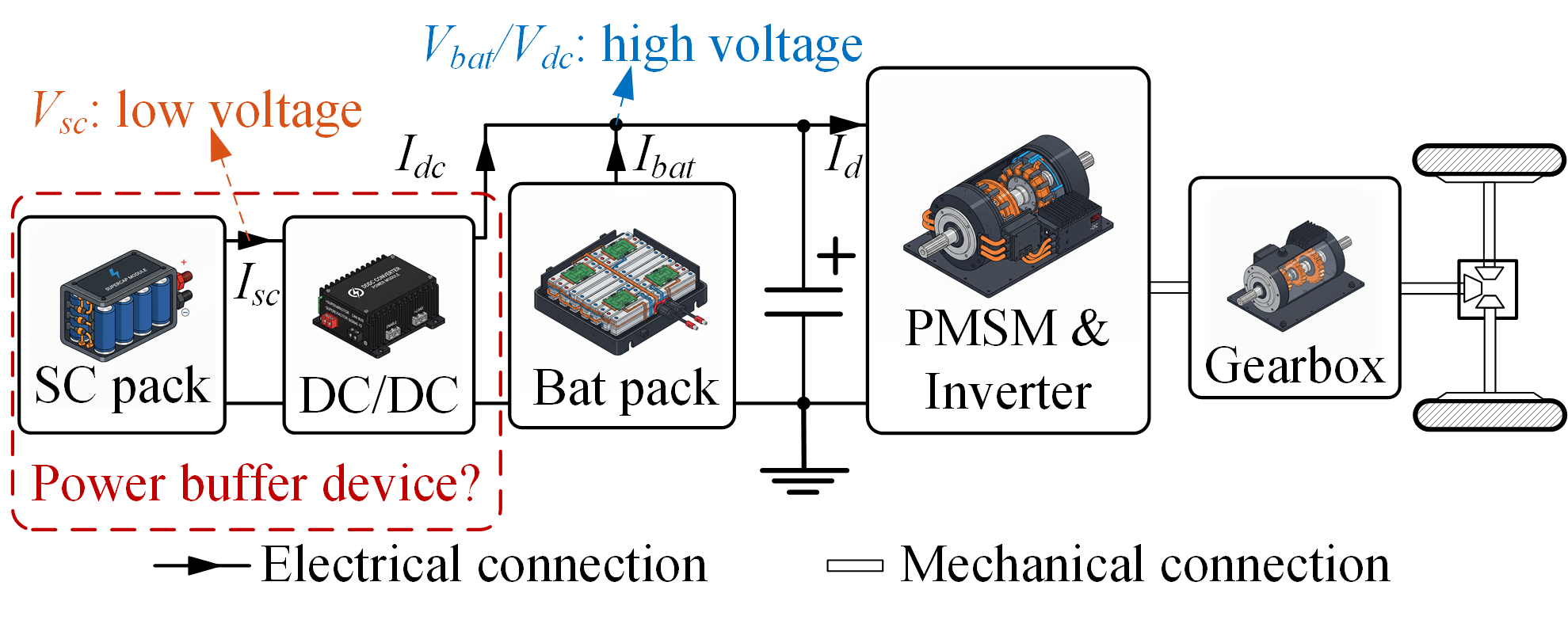} 
		\caption{Semi-active hybrid energy storage system topology for typical electric vehicle powertrain.}
		\label{fig2_config}
	\end{figure}

	The high-voltage battery pack is the sole energy source in the original battery-only EV and is capable of independently supplying the full traction power demand. Typically, it is connected to the traction motor through a DC/AC inverter for electromechanical energy conversion. Between the motor and the wheels/differentials, a single-speed reduction gearbox (and an additional AMT for the Truck) is used to increase torque and reduce rotational speed, as shown in Fig. \ref{fig2_config}.
	
	In this study, an SC pack is added to the existing EV only as an auxiliary power buffer. The semi-active topology shown in Fig. \ref{fig2_config} is adopted, where the relatively low SC voltage is boosted to the battery-pack voltage through a bidirectional DC/DC converter.
	
	\subsection{Electric vehicle model}
	
	The total wheel torque $T_{w}$ of an electric vehicle can be derived as:
	\begin{equation}
		\label{eq:Tw}
		T_{w} = (mg\sin\theta + mgC_{r}\cos\theta + \frac{1}{2} \rho A_{f} C_{d}v^2 + ma)r_{w},
	\end{equation}
	where $m$ and $g$ represent the vehicle total mass and gravitational acceleration (9.8m/s$^2$). $v$ and $a$ denote the vehicle speed and acceleration. $\theta$ is the road surface angle (in $^\circ$), which equals $\arctan(\text{slope})$. $C_{r}$ and $C_{d}$ are the rolling resistance and aerodynamic drag coefficients, respectively. $\rho$, $A_{f}$, and $r_{w}$ denote the air density, vehicle equivalent front area, and wheel radius. The longitudinal model neglects rotating-component inertia, axle-load transfer, and tire--road slip interactions.
	
	The vehicle total mass includes the vehicle mass $m_{veh}$ (with driver), load mass $m_{load}$ (e.g., passengers in the Bus and cargo in the Truck), and supercapacitor pack mass $m_{sc}$ (including packaging and management system):
	\begin{equation}
		\label{eq:mass}
		m = m_{veh} + m_{load} + m_{sc}.
	\end{equation}
	
	The relationships between the wheel torque/speed ($T_{w}$/$v$) and motor torque/rotational speed ($T_{m}$/$n_{m}$) are given as follows:
	\begin{equation}
		\label{eq:T}
			T_{m} = \frac{T_{w}}{i_{0}i_{1}},
			n_{m} = \frac{60i_{0}i_{1}v}{2 \pi r_{w}},
	\end{equation}
	where $i_{0}$ is the reduction ratio of the gearbox, $i_{1}$ is the reduction ratio of AMT (if applicable).
	
	The power demand of the motor and on the DC side can then be determined by:
	\begin{equation}
	\begin{aligned}
		P_{m} = & T_{m} \cdot \frac{2\pi n_{m}}{60} \cdot \eta_{m}^\text{tag}(T_{m}, n_{m}),\\
		P_{d} = & P_{m} \cdot \eta_{dcac}^\text{tag} =  T_{m} \cdot \frac{2\pi n_{m}}{60} \cdot \eta_{m}^\text{tag}(T_{m}, n_{m}) \cdot \eta_{dcac}^\text{tag},
	\end{aligned}
	\end{equation}
	where tag = -1 for traction and 1 for regenerative braking. $\eta_{dcac}$ and $\eta_{m}$ are the efficiencies of the DC/AC inverter and motor, respectively. In this study, $\eta_{dcac}$ is set as a constant 0.93 to represent an average operation value. $\eta_{m}$ is a look-up table of $T_{m}$ and $n_{m}$, see Supplementary Materials.
	
	\subsection{HESS electrical model}
	
	The DC-side power demand must be supplied by the HESS and therefore satisfies:
	\begin{equation}
		\begin{aligned}
			\label{eq:powerbalance}
			P_{d} = 
			\begin{cases}
				P_{bat} + P_{sc}\eta_{dcdc},  & \text{$P_{sc} \geq 0$}\\
				P_{bat} + P_{sc}/\eta_{dcdc},  & \text{$P_{sc} < 0$}
			\end{cases}
		\end{aligned}
	\end{equation}
	where $P_{bat}$ and $P_{sc}$ are the input/output power of the battery and supercapacitor packs, respectively. $\eta_{dcdc}$ is the efficiency of the DC/DC converter, an average constant 0.95 in this study.
	
	$P_{bat}$ and $P_{sc}$ are controlled by the total supercapacitor power $P_{sc,total}$ (including power losses on the internal resistances), which is also controlled through the DC/DC converter (duty cycle). The state-of-charge (SoC) of the battery pack SoC$_{bat}$ and supercapacitor pack SoC$_{sc}$, and current $I_{bat}$/$I_{sc}$ can then be derived \cite{wu2024integrated}.
	
	Parameters of the LiFePO$_4$ battery, supercapacitor, and four different electric vehicles are provided in Table \ref{tab2} and Table \ref{tab3}. The LiFePO$_4$ battery is considered the BYD Blade battery \cite{BYDblade}, and the supercapacitor is the Maxwell BCAP3000 \cite{maxwell}. Vehicle parameters (especially battery configuration) are referenced from the BYD Qin (sedan), TangL (SUV), B10 (bus), and T31 (dump truck), and have undergone certain reasonable modifications.
	
	\begin{table}[!b] 
		\caption{Parameters of the considered LiFePO$_4$ battery cell and Maxwell supercapacitor cell.}
		\label{tab2}
		\centering
		\begin{threeparttable}
			\resizebox{8cm}{!}{
				\begin{tabular}{lllcc}
					\toprule
					& Parameter & Symbol & Unit & Value  \\
					\midrule
					& Nominal voltage & $V_{bat,oc,cell}$ & V & 3.2   \\
					& Voltage range & $V_{bat,cell}$ & V & [2.5, 3.6]  \\
					& SoC range & SoC$_{bat}$ & - & [0, 1]  \\
					\multirow{2}{*}{Battery} & Nominal capacity & $Q_{bat,cell}$ & Ah & 150/119/105  \\
					& Stored energy  & $E_{bat,cell}$ & Wh & 480/380.8/336 \\
					& Internal resistance &  $r_{bat,cell}$ & m$\Omega$ & $\sim$8.2 \\
					& Weight &  $m_{bat,cell}$ & kg & 3/2.38/2.1 \\
					& Volume & Vol$_{bat,cell}$ & m$^3$ & 13.97/11.09/9.79 $\times$10$^{-4}$  \\
					\midrule
					& Nominal voltage & $V_{sc,cell,max}$ & V  & 2.7  \\
					& Voltage range & $V_{sc,cell}$ & V & [1.35, 2.7]  \\
					& SoC range & SoC$_{sc}$ & - & [0.5, 1]  \\
					\multirow{2}{*}{Supercapacitor} & Nominal capacitance & $C_{sc,cell}$ & F & 3000  \\ 
					& Stored energy & $E_{sc,cell}$ & Wh & 3.3 \\
					& Internal resistance &  $r_{sc,cell}$ & m$\Omega$ & $\sim$0.15 \\
					& Weight & $m_{sc,cell}$ & kg & 0.475 \\
					& Volume & Vol$_{sc,cell}$ & m$^3$ & 5.32$\times$10$^{-4}$ \\
					\bottomrule
			\end{tabular}}
		\end{threeparttable}
	\end{table}
	
	\begin{table*}[!t] 
		\caption{Parameters of the considered four different electric vehicles.}
		\label{tab3}
		\centering
		\begin{threeparttable}
			\resizebox{15cm}{!}{
				\begin{tabular}{llcccccc}
					\toprule
					Module & {Parameter} & Symbol & Unit & Sedan & SUV & Bus & Truck  \\
					\midrule
					& Vehicle weight without HESS & $m_{veh}$ & kg & 1800 & 2675 & 11800 & 20300 \\
					& Additional load weight & $m_{load}$ & kg & - & - & 2450$^1$ & 0/9000/18000$^2$  \\
					& Rolling resistance coefficient & $C_r$ & - & 0.01 & 0.01 & 0.007 & 0.02$^3$  \\
					& Aerodynamic drag coefficient & $C_{d}$ & - & 0.26 & 0.262 & 0.44 & 0.5  \\
					& Air density at 25 degrees & $\rho$ & kg/m$^3$ & 1.18 & 1.18 & 1.18 & 1.18  \\
					Vehicle & Frontal area & $A_{f}$ & m$^2$ & 2.2 & 2.8 & 7.2 & 7.66  \\
					& Wheel radius & $r_w$ & m & 0.34 & 0.39 & 0.46 & 0.53  \\
					& AMT reduction ratio & $i_1$ & - & - & - & - & [3.5, 3, 2.5, 1.6, 1.238, 1]  \\
					& Gear reduction ratio & $i_0$ & - & 9.34 & 10.90 & 10.2 & 6.73  \\
					& Road surface angle $^4$ & $\theta$ & $^\circ$ & 0 & 0 & 0 & 0 / variable \\
					& \multirow{2}{*}{Trunk/Cargo volume $^5$} & \multirow{2}{*}{Vol$_{cargo}$} & \multirow{2}{*}{m$^3$} & \multirow{2}{*}{0.49} & 0.2-0.26(7 seats) & \multirow{2}{*}{2.5-4.0} & \multirow{2}{*}{20} \\
					&  &  &  &  & 0.55-0.65(5 seats) &  &  \\
					\midrule
					& Motor maximum torque $^6$ & $T_{max}$ & Nm & 330 & 420 & 1960 & 3200 \\
					Motor & Motor maximum speed & $n_{max}$ & rpm & 13122 & 17800 & 4000 & 3600 \\
					& Vehicle maximum speed & $v_{max}$ & kph & 180 & 240 & 69 & 89 \\
					\midrule
					& Battery pack capacity & $Q_{bat}$ & Ah & 150 & 119 & 525 & 840 \\
					& Battery pack nominal voltage & $V_{bat,oc}$ & V & 377.6 & 844.8 & 576 & 768 \\
					Battery & Battery pack configuration & - & - & 118s 1p & 264s 1p & 180s 5p & 240s 8p \\
					& Battery pack volume $^7$ & Vol$_{bat}$ & m$^3$ & 0.220 & 0.390 & 1.175 & 2.506 \\
					& Battery pack mass $^7$ & $m_{bat}$ & kg & 414.18 & 741.42 & 2551.50 & 5846.40 \\
					\bottomrule
			\end{tabular}}
			\begin{tablenotes}
				\footnotesize
				\item[1] 35 passengers (half load) considered. $^2$ Cargo for the truck, varying with the driving cycles. $^3$ Gravel road surface.
				\item[4] $\theta = \arctan(\text{slope})$, slope has a unit of \%. For Sedan/SUV/Bus, a flat road is considered for normal urban/highway scenarios. \\ Flat road and uphill/downhill scenarios are considered for the truck. 
				\item[5] trunk for the Sedan/SUV, with the rear seats upright. Variable for the bus based on rear- or bottom-mounted battery pack. 
				\item[6] multi motor (for high-performance SUV, bus, or truck) is not considered; the value is the maximum motor torque in total.
				\item[7] including packaging and battery/thermal management system.
			\end{tablenotes}
		\end{threeparttable}
	\end{table*}
	
%
%
%
%
%
	
	Note that the volumes of the LFP battery cell and SC cell in Table \ref{tab2} are calculated from the length/width/height and diameter/height \cite{BYDblade, maxwell}. The capacity $Q_{bat}$, nominal voltage $V_{bat,oc}$, volume $\text{Vol}_{bat}$, and mass $m_{bat}$ of the battery packs in Table \ref{tab3} are calculated by:
	\begin{equation}
		\begin{cases}
			\label{eq:bat}
			Q_{bat} = Q_{bat,cell} \cdot N_{p,bat}, \\
			V_{bat,oc} = V_{bat,oc,cell} \cdot N_{s,bat}, \\
			R_{bat} = R_{bat,cell} \cdot N_{s,bat} / N_{p,bat} , \\
			\text{Vol}_{bat} = N_{s,bat} \cdot N_{p,bat} \cdot \text{Vol}_{bat,cell} / k_{vol,bat}, \\
			m_{bat} = N_{s,bat} \cdot N_{p,bat} \cdot m_{bat,cell} \cdot k_{m,bat},
		\end{cases}
	\end{equation}
	where $N_{s,bat}$ and $N_{p,bat}$ are the serial and parallel numbers of the battery cell. $\text{Vol}_{bat,cell}$ and $m_{bat,cell}$ are the volume and mass of the battery cell. $k_{vol,bat}$ and $k_{m,bat}$ are the volume and mass coefficients for battery packaging. $k_{vol,bat}$ is 0.75, $k_{m,bat}$ is 1.17/1.18/1.35/1.45 for Sedan/SUV/Bus/Truck.
	
	Similarly, for the supercapacitor pack, the maximum voltage, capacitance, volume, and mass can be derived by:
	\begin{equation}
		\begin{cases}
			\label{eq:sc}
			V_{sc,max} = V_{sc,cell,max} \cdot N_{s,sc}, \\
			C_{sc} = C_{sc,cell} \cdot N_{p,sc} / N_{s,sc}, \\
			\text{Vol}_{sc} = N_{s,sc} \cdot N_{p,sc} \cdot \text{Vol}_{sc,cell} \cdot k_{vol,sc}, \\
			m_{sc} = N_{s,sc} \cdot N_{p,sc} \cdot m_{sc,cell} \cdot k_{m,sc},
		\end{cases}
	\end{equation}
	where $N_{s,sc}$ and $N_{p,sc}$ are the serial and parallel numbers of the supercapacitor cell.
	
	In this study, the volume and mass coefficients for supercapacitor packaging, $k_{vol,sc}$ and $k_{m,sc}$, are set to 1.25 and 1.15, respectively. Subscript $cell$ indicates variables for the battery or supercapacitor cell; no subscript $cell$ means battery or supercapacitor pack. 
	
	\subsection{Capacity loss model for the battery pack}
	
	A recognized battery aging model for LiFePO$_4$ battery \cite{wang2011cycle} is adopted in this study:
	\begin{equation}
		\label{eq:batlife1}
		Q_{loss} = A \cdot e^{-(\frac{E_{a} + B \cdot C_{rate}}{RT_{bat}})} \cdot A^{z}_{h},
	\end{equation}
	which has been experimentally validated and calibrated by Song et al. \cite{song2015optimization}. Its discrete form  is listed below:
	\begin{footnotesize}
		\begin{equation}	
		\begin{aligned}
			\label{eq:Qlossk}
			\Delta Q_{loss,t} & = \Delta A_{h} z A^{\frac{1}{z}} e^{-(\frac{E_{a}+B \cdot C_{rate,t}}{z RT_{bat}})} Q_{loss,t-1}^{\frac{z-1}{z}} \\
			& = 9.78 \times 10^{-4} \Delta A_{h} e^{(\frac{-15162+1516C_{rate,t}}{0.849RT_{bat,t}})} Q_{loss,t-1}^{-0.1779},
		\end{aligned}
		\end{equation}
	\end{footnotesize}
	where $\Delta Q_{loss,t}$ indicates the capacity loss at time instant $t$, $R$ denotes the gas constant which equals 8.314J/(mol$\cdot$K). $T_{bat,t}$ is the absolute battery temperature, 298K in this study (temperature is constant under short distance driving \cite{wu2024optimal}). $C_{rate,t}$ represents the battery C-rate (i.e., $|I_{bat,t}|/Q_{bat}$). $\Delta A_{h}$ is the Ah-throughput.
	
	The above battery capacity loss model is for a battery cell, thus $\Delta A_{h} = \frac{|I_{bat}|T_{s}}{3600}$ \cite{song2015optimization}, where $T_s$ is the sampling time, 1s. In this study, the variable $I_{bat}$ denotes the battery pack current (without subscript $cell$). Thus, the battery capacity loss model for the battery pack should be calculated very carefully together with Eq. (\ref{eq:bat}):
	\begin{footnotesize}
		\begin{equation}	
			\label{eq:Qloss}
			\Delta Q_{loss,t} = 9.78 \times 10^{-4} \frac{\bm{\frac{|I_{bat,t}|}{N_{p,bat}}}T_{s}}{3600} e^{(\frac{-15162+1516 \frac{|I_{bat,t}|}{Q_{bat}}}{0.849RT_{bat,t}})} Q_{loss,t-1}^{-0.1779}.
		\end{equation}
	\end{footnotesize}
	
	This model needs to be highlighted. For battery packs with a multiple-parallel configuration, such as the 5p/8p for the bus/truck in Table \ref{tab3}, directly using Eq. (\ref{eq:Qlossk}) can result in unreasonable rapid battery degradation (5x/8x), leading to higher operation costs afterward.
	
	Note that the model in Eq. (\ref{eq:Qloss}) is general for battery packs under the following assumptions:
	
	\textbf{(1) Thermal assumption}: The battery pack is thermally uniform, i.e., each cell has the same temperature.
	
	\textbf{(2) Equilibrium assumption}: The pack current can be allocated to each battery in parallel, on average.

	\section{Offline supercapacitor sizing and online energy management strategy implementation \label{method}}
	
	In this section, the detailed process for offline supercapacitor sizing optimization is introduced. In particular, the determination of the feasible supercapacitor serial-parallel candidate is described in detail, considering pack voltage, mass, and volume constraints. The optimal energy management using dynamic programming is provided. Finally, online energy management using DRL-BC is presented.
	
	\subsection{Supercapacitor serial candidates determination based on pack voltage}
	
	\begin{figure}[!b]
		\centering
		\includegraphics[width=8cm]{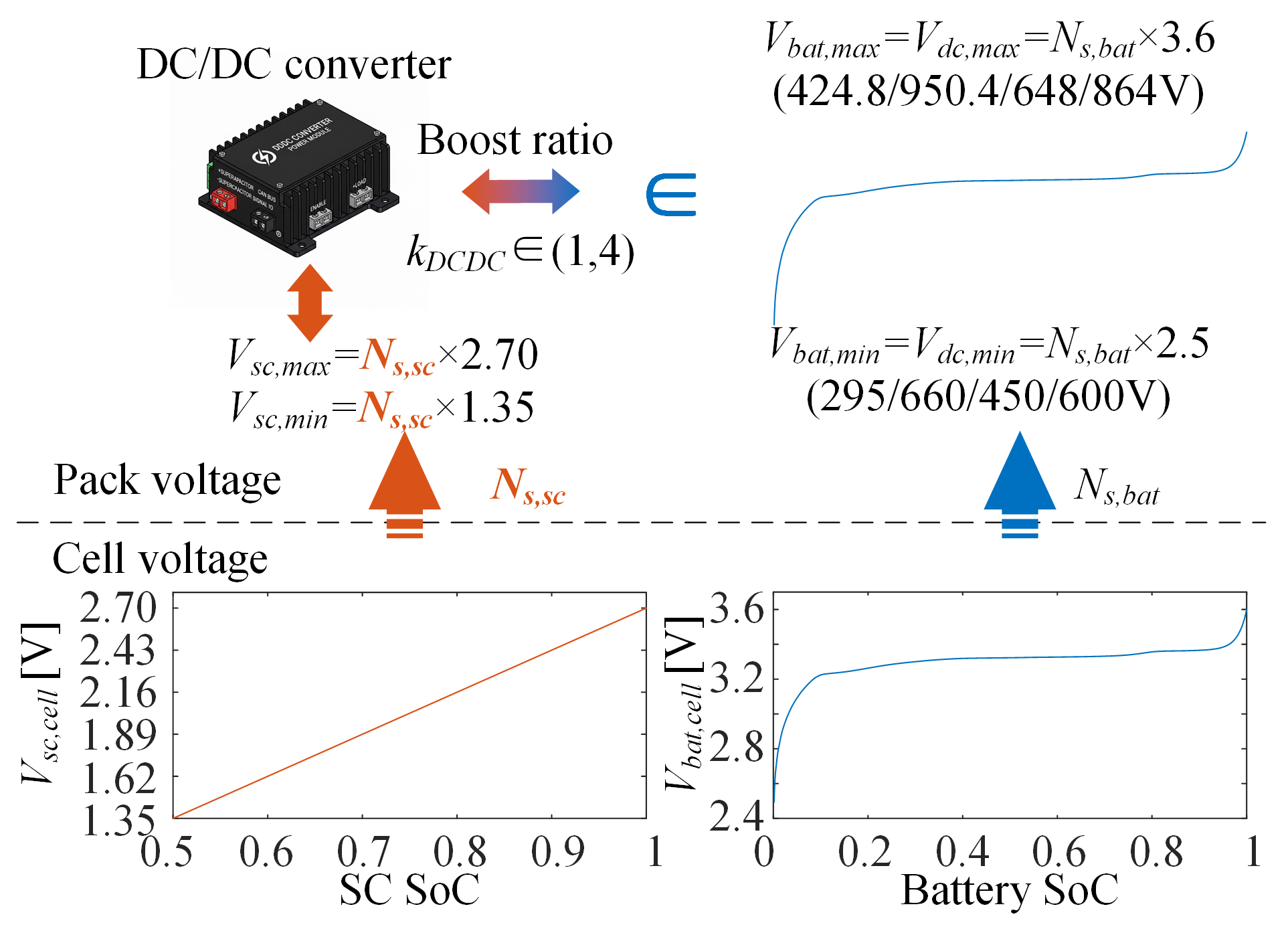}
		\caption{SC series number determination. 424.8/950.4/648/864V are the maximum battery pack voltages of the Sedan/SUV/Bus/Truck, 295/660/450/600V are the minimum battery pack voltages.} 
		\label{fig3_SC_config}
	\end{figure}
	
	The DC-DC converter shown in Fig. \ref{fig2_config} must be capable of boosting the supercapacitor pack voltage to match the LFP battery pack voltage, across the entire SoC ranges of both energy storage units, as depicted in Fig. \ref{fig3_SC_config}. Therefore, the system must satisfy the following voltage boundary conditions:
	\begin{equation}
		\begin{cases}
			\label{eq:Vsc_pack}
			N_{s,sc}V_{sc,cell,min}k_{DCDC,max}  \geq V_{bat,pack,max}  \\
			N_{s,sc}V_{sc,cell,max}k_{DCDC,min}  \leq V_{bat,pack,min}
		\end{cases}
	\end{equation}
	where $k_{DCDC,max}$ is the maximum voltage boost ratio of the DC-DC converter, set to 4, and $k_{DCDC,min}$ is the minimum boost ratio, equal to 1. The first inequality ensures that even at the minimum supercapacitor pack voltage, the converter can boost it to the maximum battery pack voltage. The second inequality dictates that the maximum supercapacitor pack voltage must not exceed the minimum battery pack voltage, which prevents uncontrolled power flow and ensures proper converter operation. Thus, the allowable range for the number of series-connected supercapacitors is derived as:
	\begin{equation}
		\label{eq:sc_serial}
		\frac{V_{bat,pack,max}}{k_{DCDC,max} V_{sc,cell,min}} \leq N_{s,sc}  \leq \frac{V_{bat,pack,min}}{k_{DCDC,min} V_{sc,cell,max}}.
	\end{equation}
	
	Based on the vehicle configurations detailed in Table \ref{tab3}, the feasible number of series-connected supercapacitors for the Sedan, SUV, Bus, and Truck should fall within the ranges of [79, 109], [176, 244], [120, 166], and [160, 222], respectively.
	
	\subsection{Supercapacitor parallel candidates determination based on mass/volume}
	
	\begin{figure*}[!h]
		\centering
		\includegraphics[width=16cm]{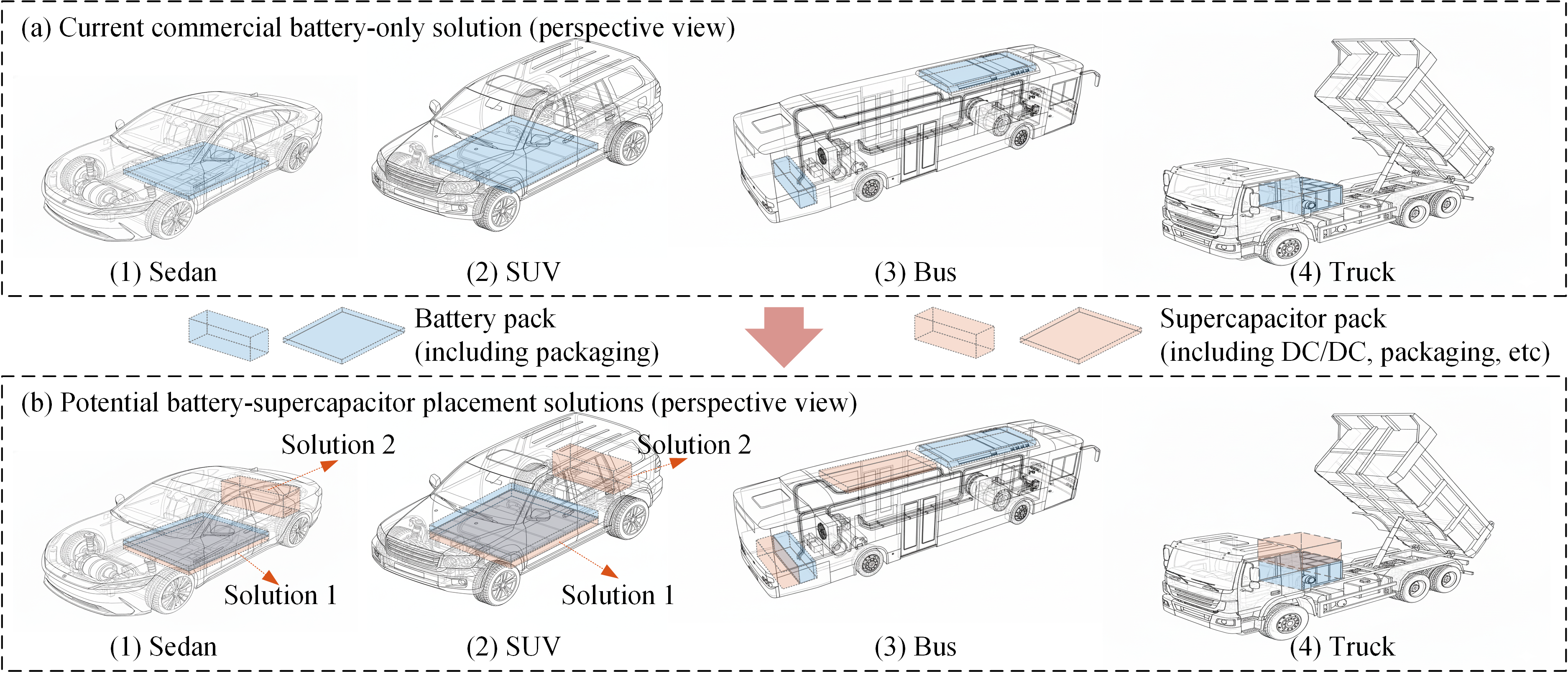}
		\caption{Perspective views of the (a) current commercial battery-only placement solution and the (b) potential battery-supercapacitor placement solutions.}
		\label{fig4_perspective_view}
	\end{figure*}
	
	\begin{figure*}[!h]
		\centering
		\includegraphics[width=16cm]{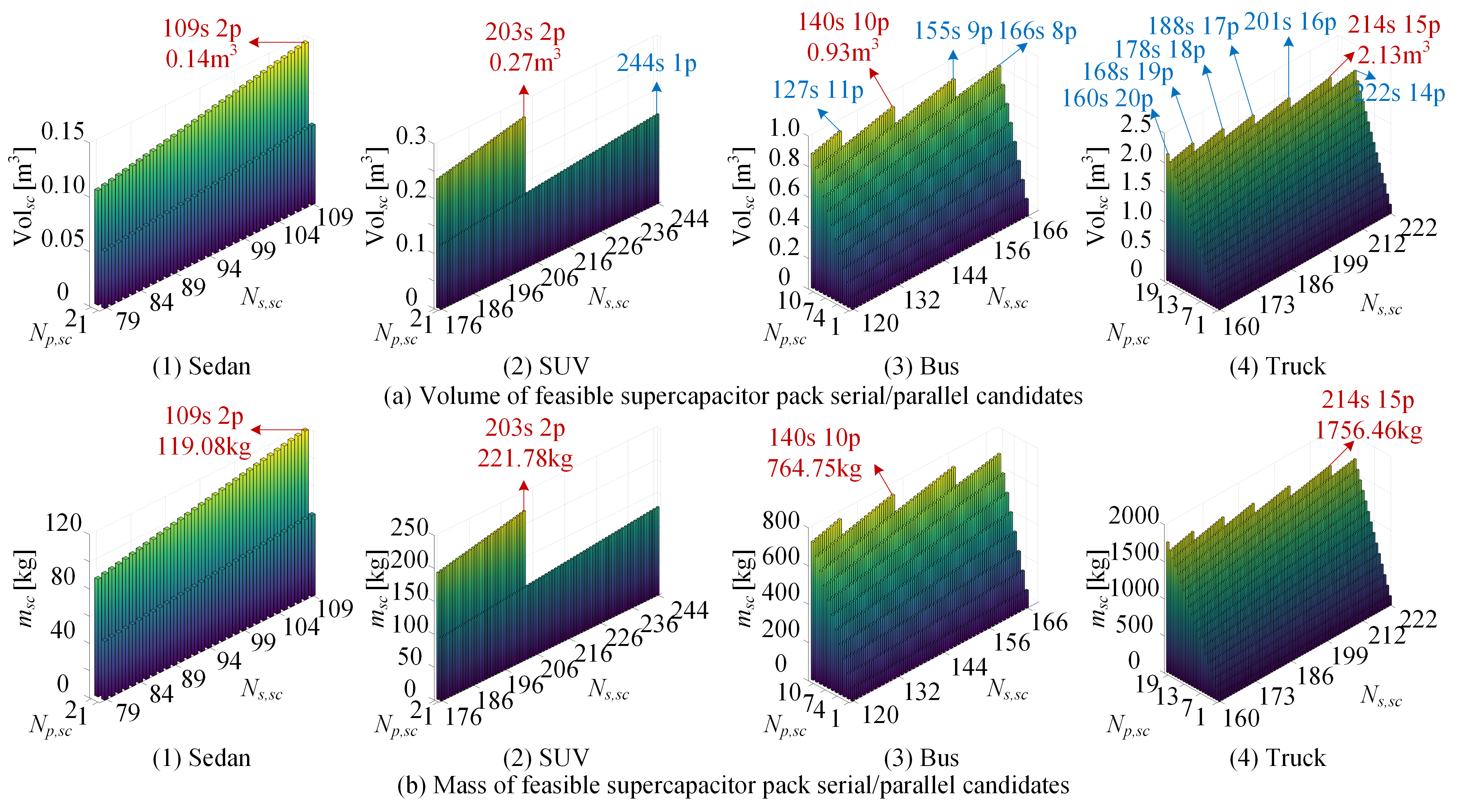}
		\caption{(a) Volume and (b) Mass of feasible supercapacitor serial/parallel candidates for the Sedan, SUV, Bus, and Truck. The heaviest/biggest candidate is marked using red color.} 
		\label{fig5_mass_volume}
	\end{figure*}
	
	To determine the number of supercapacitor in parallel, the added mass and volume must be constrained. In this study, the additional supercapacitor pack mass cannot exceed 30\% of the battery pack mass to avoid a significant increase in load on electric vehicles. Meanwhile, the additional supercapacitor pack volume cannot exceed 100\% of the battery pack.
	
	The 30\% mass ratio is introduced as a conservative engineering screening limit rather than a universal industrial standard. Additional energy storage mass increases the traction power demand and may alter the optimal HESS sizing and operating cost \cite{huang2022sizing}. Meanwhile, electric vehicles inherently	retain a load-carrying margin for passengers or cargo. For the four platforms considered here, the adopted upper limit corresponds to only approximately 6.5--8.6\% of the base vehicle mass and is substantially below the operational loads considered.
	
	A different percentage limit is adopted for volume due to the energy density difference between the battery and SC. After considering the packaging coefficients, the volumetric energy densities of the battery and SC packs are about 258 and 5~Wh/L, respectively, while their mass energy densities are about 110--137 and 6~Wh/kg. The 100\% battery pack volume is used as the maximum outer geometric boundary for candidate selection.
	
	Then, for each feasible supercapacitor serial candidate $N_{s,sc,i}$, there is a distinct maximum parallel number $N_{p,sc,max}$ that can be determined:
	\begin{equation}
		\begin{cases}
			\label{eq:sc_parallel}
			N_{p,sc,max,mass} = \text{floor}(\frac{0.3 m_{bat}}{N_{s,sc,i} m_{sc,cell} k_{m,sc}})  \\
			N_{p,sc,max,vol} = \text{floor}(\frac{\text{Vol}_{bat}}{N_{s,sc,i} \text{Vol}_{sc,cell} k_{vol,sc}}) \\
			N_{p,sc,max} = \text{min}(N_{p,sc,max,mass}, N_{p,sc,max,vol})
		\end{cases}
	\end{equation}
	where $\text{floor}(\cdot)$ is the round down operation, $N_{p,sc,max,mass}$ and $N_{p,sc,max,vol}$ are the maximum supercapacitor parallel number restricted by mass and volume, respectively. Other variables are provided in Eq. (\ref{eq:bat}) - (\ref{eq:sc}). Then, the feasible supercapacitor parallel number should belong to [1, $N_{p,sc,max}$]. Note that once $N_{s,sc,i}$ is changed, $N_{p,sc,max}$ must be recalculated.
	
	In this study, potential supercapacitor placement solutions are provided for the Sedan/SUV/Bus/Truck, as illustrated in Fig. \ref{fig4_perspective_view}. The original battery pack is highlighted using light blue color in the perspective views, see Fig. \ref{fig4_perspective_view} (a). Suppose a supercapacitor pack with the same volume should be placed in the vehicle, potential solutions are provided in Fig. \ref{fig4_perspective_view} (b). For the Sedan and SUV, the additional supercapacitor pack can be placed at the chassis (Solution 1) or in the trunk (Solution 2). The trade-off is that the cabin would need to be raised, or trunk space would be reduced or even eliminated; see the trunk volume of Sedan/SUV in Table \ref{tab3}. For the Bus and Truck, there will be much more available space to place the supercapacitor pack, e.g., the roof of the Bus, the space between the cabin and the cargo hold of the Truck.
	
	Note that the bidirectional DC/DC converter is also an essential component of the HESS, and its rated power varies considerably among the investigated vehicle types. In this study, the rated powers of the converter for Sedan/SUV/Bus/Truck are set as 50kW/50kW/100kW/300kW, respectively. To maintain a consistent technological and packaging boundary, the 100kW ADVANTICS ADB-PC-DC01 \cite{DCDC100kW} bidirectional isolated DC/DC converter	was adopted as the reference power module. One module was assigned to the 50 and 100kW applications, whereas three modules connected in parallel were considered for the 300kW application. Accordingly, the converter-module masses were estimated as 47, 47, and 141kg \cite{DCDC100kW}, while the corresponding envelope volumes were 0.0462, 0.0462, and 0.1386m$^3$, respectively. These estimates exclude external filters, contactors, cooling manifolds, busbars, and installation clearances and therefore represent lower-bound packaging requirements. Compared with the maximum potential SC pack in Fig. \ref{fig4_perspective_view}, the mass and volume of the DC/DC converter are very low. Moreover, the optimal SC pack is significantly lighter/smaller than the maximum value, which will be presented in the following section. Thus, the DC/DC converter is not plotted in Fig. \ref{fig4_perspective_view}.
	
	\begin{figure*}[!t]
		\centering
		\includegraphics[width=16cm]{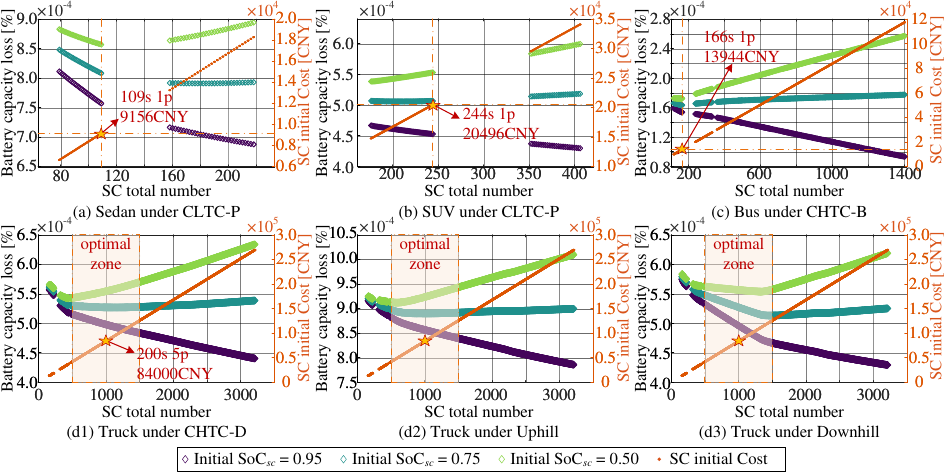}
		\caption{Battery capacity loss and supercapacitor initial cost with different supercapacitor sizing candidates for (a) Sedan under CLTC-P, (b) SUV under CLTC-P, (c) Bus under CHTC-B, and (d) Truck under (d1) CHTC-D, (d2) Uphill cycle, and (d3) Downhill cycle. The supercapacitor total number is set as the x-axis without serial/parallel info. The interrupt parts of the rhombus are infeasible candidates. The optimal sizing results are highlighted.} 
		\label{fig6_optimalsizing}
	\end{figure*}
	
	All feasible supercapacitor serial/parallel candidates for the considered Sedan, SUV, Bus, and Truck are illustrated in Fig. \ref{fig5_mass_volume}; the heaviest candidate is marked in red. Compared with existing sizing studies \cite{huang2022sizing}, the selection of feasible candidates is more meticulous. The total number of candidates is reduced, and the time wasted on DP optimization for infeasible candidates is avoided.
	
	\subsection{Dynamic programming for global optimal energy management}
	
	DP is adopted for offline global-optimal energy management for each feasible supercapacitor serial/parallel candidate for each vehicle type, aiming to determine the optimal sizing. 
	
	The SoC$_{sc}$ state space is discretized into 100 intervals, and the control variable $P_{sc,total}$ is discretized into 251 intervals. The final SoC$_{sc}$ is set to 0.75 to maintain a middle level for future acceleration/regenerative braking.
	
	\begin{figure}[!b]
		\centering
		\includegraphics[width=8cm]{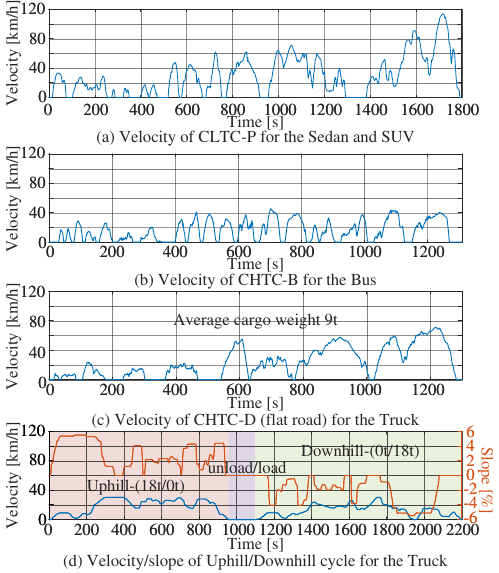}
		\caption{Four test velocity profiles for four different electric vehicles. Uphill cycle in (d) means 18t uphill and 0t downhill, Downhill cycle in (d) means 0t uphill and 18t downhill.}
		\label{fig7_testSpeed}
	\end{figure}
	
	The objective function of the DP energy management is as follows:
	\begin{equation}
		\footnotesize
		\begin{aligned}
			\label{eq:DP}
			J_{DP} = \sum_{k=1}^{k_{end}}(\frac{Q_{bat}V_{bat,oc}price_{bat}}{1000} \times \frac{\Delta Q_{loss,t}}{0.2} + \frac{price_{ele}(P_{bat,t}+P_{sc,t})}{1000 \times 3600})T_{s},
		\end{aligned}
	\end{equation}
	where $\Delta Q_{loss,t}$ is calculated from Eq. (\ref{eq:Qloss}). $price_{bat}$ is the battery price, 108USD/kWh, i.e., 756CNY/kWh (exchange rate of 7 in this study), according to the newest data from U.S. Department of Energy and BloombergNEF \cite{BatSCprice,Batprice1}. $price_{ele}$ is the Chinese commercial electricity price, 0.6CNY/kWh. Note that for each candidate, the vehicle mass and load power are updated, and DP is executed once. Details of DP can be found in \cite{song2015optimization, wu2024optimal}.
	
	\subsection{Optimal sizing}
	
	DP results for the four different electric vehicles on battery capacity loss are provided in Fig. \ref{fig6_optimalsizing}. Sedan and SUV are tested under CLTC-P; the Bus is tested under the CHTC-B driving cycle; and the Truck is tested under the CHTC-D, Uphill, and Downhill driving cycles. These dedicated driving cycles are shown in Fig. \ref{fig7_testSpeed}. For the Bus under CHTC-B and the Truck under CHTC-D, half loads of 2450kg and 9000kg are considered. Specifically, for the Truck, transportation tasks at mine shafts or mines are also considered, i.e., Uphill and Downhill cycles \cite{tang2024optimal, yang2019implementation} shown in Fig. \ref{fig7_testSpeed} (d).
	
	In Fig. \ref{fig6_optimalsizing} (a), it is clear that the 1p candidates and 2p candidates are separated. And the separated candidates can be smoothly connected by the infeasible ones. For an initial SoC$_{sc}$ of 0.95, the more supercapacitor cells, the lower battery capacity loss due to enough stored energy. When the initial SoC$_{sc}$ is 0.75, the same as the final value, there is a minimum battery degradation point between 160 and 180 cells, which means more supercapacitor cells can be a burden; the battery pack must output more energy to maintain SoC$_{sc}$. For an initial SoC$_{sc}$ of 0.50, the minimum battery degradation point comes earlier. By considering the three different initial SoC$_{sc}$ values, the optimal supercapacitor sizing is 109s 1p, marked in red. The current supercapacitor price is still high, 2000-6000USD/kWh from EIT InnoEnergy \cite{SCprice1}. In this study, an average of 4000USD/kWh, i.e., 28000CNY/kWh, is adopted. Thus, the optimal supercapacitor pack will cost about 9156 CNY.
	
	\begin{table}[!b] 
		\caption{Summary of optimal supercapacitor sizing for different vehicles.}
		\label{tab4}
		\centering
		\begin{threeparttable}
			\resizebox{8cm}{!}{
				\begin{tabular}{lllcc}
					\toprule
					Vehicle & Parameter & Symbol & Unit & Value  \\
					\midrule
					& Max pack voltage & $V_{sc,max}$ & V & 294.3   \\
					& Min pack voltage & $V_{sc,min}$ & V & 147.2  \\
					& Configuration & $N_{s,sc}$ $N_{p,sc}$ & - & 109s 1p  \\
					Sedan & Nominal capacitance & $C_{sc}$ & F & 27.52  \\
					& Stored energy & $E_{sc}$ & kWh & 0.3597 \\
					& Pack mass & $m_{sc}$ & kg & 59.54 \\
					& Pack volume & Vol$_{sc}$ & m$^3$ & 0.072 \\
					\midrule
					& Max pack voltage & $V_{sc,max}$ & V &  658.8  \\
					& Min pack voltage & $V_{sc,min}$ & V & 329.4  \\
					& Configuration & $N_{s,sc}$ $N_{p,sc}$ & - & 244s 1p  \\
					SUV & Nominal capacitance & $C_{sc}$ & F & 12.30  \\
					& Stored energy & $E_{sc}$ & kWh & 0.8052 \\
					& Pack mass & $m_{sc}$ & kg & 133.28 \\
					& Pack volume & Vol$_{sc}$ & m$^3$ & 0.162 \\
					\midrule
					& Max pack voltage & $V_{sc,max}$ & V & 448.2   \\
					& Min pack voltage & $V_{sc,min}$ & V & 224.1  \\
					& Configuration & $N_{s,sc}$ $N_{p,sc}$ & - & 166s 1p  \\
					Bus & Nominal capacitance & $C_{sc}$ & F & 18.07  \\
					& Stored energy & $E_{sc}$ & kWh & 0.5478 \\
					& Pack mass & $m_{sc}$ & kg & 90.68 \\
					& Pack volume & Vol$_{sc}$ & m$^3$ & 0.110 \\
					\midrule
					& Max pack voltage & $V_{sc,max}$ & V & 540.0   \\
					& Min pack voltage & $V_{sc,min}$ & V & 270.0  \\
					& Configuration & $N_{s,sc}$ $N_{p,sc}$ & - & 200s 5p  \\
					Truck & Nominal capacitance & $C_{sc}$ & F & 75.00  \\
					& Stored energy & $E_{sc}$ & kWh & 3.3 \\
					& Pack mass & $m_{sc}$ & kg & 546.25 \\
					& Pack volume & Vol$_{sc}$ & m$^3$ & 0.665 \\
					\bottomrule
			\end{tabular}}
		\end{threeparttable}
	\end{table}
	
	Similarly, the SUV and Bus also have three separated but continuous battery capacity loss Pareto fronts, see Fig. \ref{fig6_optimalsizing} (b) and Fig. \ref{fig6_optimalsizing} (c). And the optimal sizing results are both selected as the maximum 1p, 244s 1p (20496CNY) and 166s 1p (13944CNY), respectively. Note that the SUV uses the latest kilovolt battery platform (about 950 V); the supercapacitor pack must also be high-voltage. Even when the initial SoC$_{sc}$ is above 0.75, it still results in a low battery capacity loss. Also, the optimal sizing of the Bus is significantly cheaper than that of the SUV , since urban bus driving cycles involve frequent starts and stops and a relatively low top velocity.
	
	Regarding the Truck, it can be seen that there are three minimum battery degradation zones (at the 0.75 curves) for CHTC-D, Uphill, and Downhill cycles in Fig. \ref{fig6_optimalsizing} (d-f), from 500 to 1500 cells. For an initial SoC$_{sc}$ of 0.95, the battery capacity loss curves are all monotonically decreasing under the three driving cycles. For an initial SoC$_{sc}$ of 0.5, the minimum battery degradation zones come earlier. Thus, the optimal sizing is selected as 200s 5p (84000CNY) to suit different driving cycles. Note that the battery capacity loss under Uphill is significantly higher than that under CHTC-D and Downhill, since the full-load uphill requires much more energy.
	
	From Fig. \ref{fig6_optimalsizing} (a-f), it can also be concluded that the optimal battery capacity loss has nothing to do with the supercapacitor serial-parallel configuration, i.e., as long as the cell total number is the same, different serial-parallel configurations will result in the same battery degradation (many rhombuses overlap). Thus, high-voltage configuration with fewer parallel number is recommended for lower line power loss. The optimal SC sizing results for Sedan/SUV/Bus/Truck are summarized in Table \ref{tab4}, along with the corresponding physical parameters. Note that for Sedan/SUV/Bus/Truck, the SC pack accounts for 14.4\%/18.0\%/3.6\%/9.3\% of the battery pack mass and 32.7\%/41.5\%/9.4\%/26.5\% of its volume, substantially below the upper limits.
	
	\subsection{Online energy management using DRL-BC}
	
	To achieve practical energy management performance through optimal sizing, a state-of-the-art online method, TD3 initialized by BC, is adopted. 
	
	The energy management strategy can be formulated as a discrete-time Markov decision process. At each time step $t$, the agent observes the environment's state $s_t$, selects an action $a_t$ according to the current policy, and subsequently receives a scalar reward $r_t$ from the environment. The core elements, namely the state vector, action vector, and reward function, are unified and defined as follows:
	\begin{equation}
		\footnotesize
		\left\{
		\begin{aligned}
			s_t &= [\widetilde{P}_{d,t}, \text{SoC}_{bat,t}, \text{SoC}_{sc,t}, \hat{\tau}_t]^T \qquad
			\widetilde{P}_{d,t} = \frac{P_{d,t}-\mu_{P_d}}{\sigma_{P_d}}, \\
			a_t &= P_{sc,total,t} \in \mathcal{A}, \\
			r_t &= \mathcal{R}(s_t,a_t). 
		\end{aligned}
		\right.
		\label{eq:mdp_def}
	\end{equation}
	where $\mathcal{S}$ and $\mathcal{A}$ denote the continuous state and action spaces, respectively. 
	$\widetilde{P}_{d,t}$ is the normalized load power, where $\mu_{P_d}$ and $\sigma_{P_d}$ are obtained from the corresponding training cycle; $\hat{\tau}_t$ denotes the estimated normalized trip progress; and $\mathcal{R}(\cdot)$ represents the reward function. The reward primarily minimizes battery degradation while regulating the terminal $\text{SoC}_{sc}$ toward its initial value of 0.75. Detailed definitions of the trip progress estimation, reward formulation, and physical-constraint handling are provided in the Supplementary Materials.
	
	\begin{figure*}[!t]
		\centering
		\includegraphics[width=16cm]{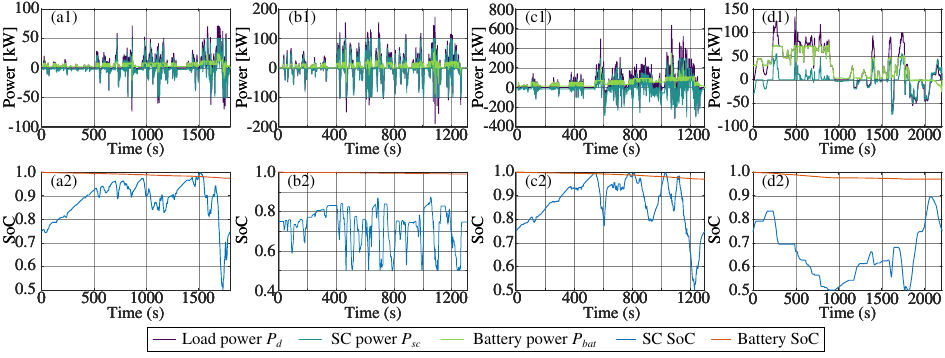}
		\caption{Optimal DP load power allocation and battery/supercapacitor SoC for (a) SUV under CLTC-P, (b) Bus under CHTC-B, (c) Truck under CHTC-D, (d) Truck under Downhill cycle. Initial SC SoC equals 0.75.} 
		\label{fig8_DPpowerallocation}
	\end{figure*}
	
	\begin{figure*}[!t]
		\centering
		\includegraphics[width=16cm]{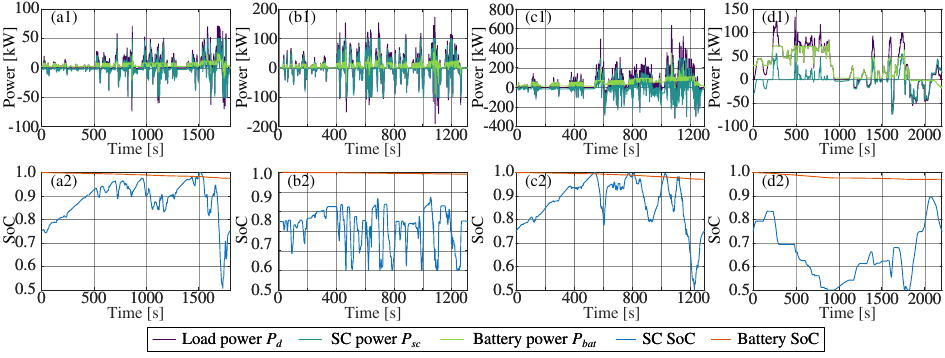}
		\caption{Online DRL-BC load power allocation and battery/supercapacitor SoC for (a) SUV under CLTC-P, (b) Bus under CHTC-B, (c) Truck under CHTC-D, (d) Truck under Downhill cycle. Initial SC SoC equals 0.75.}
		\label{fig9_DRLpowerallocation}
	\end{figure*}
	
	Note that the maximum and minimum values of action $P_{sc,total}$ are different for Sedan/SUV/Bus/Truck considering the DC/DC converter size, which are set $\pm$50kW/50kW/100kW/300kW, respectively.
	
	To introduce expert knowledge into the online controller, the DP trajectories are first used to train the Actor through behavioral cloning. The BC objective is defined as
	\begin{equation}
		\footnotesize
		\mathcal{L}_{BC}
		=
		\frac{1}{N_b}
		\sum_{i=1}^{N_b}
		\left\|
		\pi_{\phi}(s_i)-a_{DP,i}
		\right\|_2^2,
		\label{eq:bc_loss}
	\end{equation}
	where $\pi_{\phi}$ denotes the Actor policy, $a_{DP,i}$ is the corresponding DP expert action, and $N_b$ is the mini-batch size. The BC-trained Actor is subsequently used to initialize the DRL-BC policy. During online TD3 fine-tuning, the Actor objective is
	\begin{equation}
		\footnotesize
		\mathcal{L}_{actor}
		=
		-\frac{1}{N_b}
		\sum_{i=1}^{N_b}
		Q_{\theta_1}\left(s_i,\pi_{\phi}(s_i)\right)
		+
		\lambda_{BC}\mathcal{L}_{BC},
		\label{eq:drlbc_actor_loss}
	\end{equation}
	where $Q_{\theta_1}$ is the first Critic network and $\lambda_{BC}$ is a decaying expert coefficient. Therefore, conventional TD3 and DRL-BC share the same TD3 backbone and online interaction process, while DRL-BC additionally employs BC initialization and the expert-retention term in Eq. (\ref{eq:drlbc_actor_loss}). All BC, conventional TD3, and DRL-BC experiments were conducted using a fixed random seed of 42. Terminal charge sustainability was evaluated using $|\mathrm{SoC}_{sc,T}-0.75|\leq0.001$ for TD3 and DRL-BC. Detailed network architecture, training parameters, convergence criteria, and implementation settings are provided in the Supplementary Materials. More methodology details can also be found in \cite{zhang2022twin, peng2024efficient}.
	
	\begin{table*}[!t] 
		\caption{Operation cost and final SC SoC comparisons of DP, Rule, MPC, BC, TD3, and DRL-BC energy management strategies. Results are presented with the format of operation cost / final SC SoC, the operation cost has the unit of CNY/100km.}
		\label{tab_comp}
		\centering
		\begin{threeparttable}
			\resizebox{15cm}{!}{
			\begin{tabular}{lcccccc}
					\toprule
					Method & Sedan (CLTC-P) & SUV (CLTC-P) & Bus (CHTC-B) & Truck (CHTC-D) & Truck (Downhill) & Truck (Uphill) \\
					\midrule
					Bat-only & 31.87 / - & 37.18 / - & 150.85 / - & 357.61 / - & 288.46 / - & 445.48 / - \\
					DP       & 19.15 / 0.75 & 23.38 / 0.75 & 65.27 / 0.75 & 286.33 / 0.75 & 264.78 / 0.75 & 442.84 / 0.75\\
					Rule$^1$ & 27.94 / 0.85 & 27.23 / 0.73 & 67.02 / 0.77 & 313.02 / 0.74 & 289.59 / 1.00 & 453.64 / 0.91\\
					MPC$^2$  & 27.25 / 0.75 & 30.87 / 0.75 & 96.98 / 0.75 & 331.99 / 0.75 & 294.86 / 0.75 & 466.34 / 0.75\\
					BC       & 19.25 / 0.77 & 23.47 / 0.75 & 65.58 / 0.75 & 286.77 / 0.75 & 264.92 / 0.76 & 442.94 / 0.74\\
					TD3      & 23.52 / 0.75 & 28.46 / 0.75 & 80.93 / 0.75 & 310.91 / 0.75 & 269.05 / 0.75 & 458.66 / 0.75\\
					DRL-BC   & 19.24 / 0.75 & 23.40 / 0.75 & 65.43 / 0.75 & 286.44 / 0.75 & 264.88 / 0.75 & 442.97 / 0.75\\
					\bottomrule
			\end{tabular}}
			\begin{tablenotes}
				\footnotesize
				\item[1] Rules are extracted from the DP results of Sedan/SUV/Bus/Truck (CHTC-D). Results for Truck (Downhill/Uphill) are executed from the CHTC-D rules for simplicity, thus perform worse than battery-only in terms of operation cost and SoC.
				\item[2] MPC has a prediction horizon of 10(s) and perfect DC-side power demand $P_{d}$ prediction.
			\end{tablenotes}
		\end{threeparttable}
	\end{table*}
	
	The global optimal energy management results from DP are shown in Fig. \ref{fig8_DPpowerallocation}. The battery mainly provides the relatively smooth power component, whereas the SC absorbs regenerative braking energy and supplies transient peak power. The $\text{SoC}_{sc}$ is allowed to vary between 0.5 and 1.0 and is regulated toward its initial value at the end of the driving cycle. Due to space limitations, the results for the Sedan and Truck Uphill are not presented.
	
	Regarding online energy management using DRL-BC, the power allocation and $\text{SoC}_{sc}$ trajectories are close to those obtained by DP, as shown in Fig. \ref{fig9_DRLpowerallocation}. Unlike DP, which exploits the complete driving-cycle information offline, DRL-BC determines the power allocation online according to the observed state. Consequently, small deviations from the DP trajectories and terminal $\text{SoC}_{sc}$ are unavoidable. Quantitative comparisons with battery-only, DP, the DP-informed rule-based strategy \cite{song2015optimization}, MPC \cite{wang2023optimized}, BC, and conventional TD3 are listed in Table \ref{tab_comp}. Details of the rule-based/MPC/BC/conventional TD3 strategies and complete convergence/ablation results are provided in the Supplementary Materials. Frozen policies were additionally evaluated without retraining on unseen driving cycles, including WLTP for the Sedan/SUV and alternative CHTC cycles for the Bus and Truck. No unseen-cycle data were used for policy training, normalization recalibration, hyperparameter adjustment, or checkpoint selection. Note that the DRL-BC is very close to DP thanks to the $\tau_t$ state and the BC expert, see Figs. S2-S6 and Table S9. Complete frozen-policy results on unseen driving cycles are reported in Table S6, showing that distribution shifts can affect both battery degradation and SC SoC regulation.
	
	The term DRL-BC is used here to emphasize that BC only provides the expert initialization and regularization, followed by online reinforcement-learning interaction and policy optimization. This differs from TD3-BC formulations in which the learned policy is trained primarily or entirely from an offline dataset. The DRL-BC implementation in this study is intended as a representative near-optimal online EMS rather than the central methodological contribution of the overall feasibility framework. Other advanced online energy management strategies, such as safe reinforcement learning \cite{li2026safe,jia2026decoupled}, offline reinforcement learning \cite{wang2025data}, and integration with driving pattern \cite{li2025adaptive}, can also be incorporated into the proposed framework.
	
	The physical power flow is continuous, whereas the DP and DRL-BC power commands are updated at a discrete 1-s control interval. Converter switching dynamics are assumed to be sufficiently faster than the EMS timescale and are therefore not modeled explicitly. After training, online implementation only requires a forward inference of the Actor network to determine the SC power. Critic-network evaluation, replay-buffer updating, and gradient-based optimization are not required during onboard deployment, substantially reducing the real-time computational burden compared with the training stage.

	\section{Economy and feasibility analysis \label{result1}}
	
	This section includes four parts: (1) the overall economy analysis and comparison with battery-only electric vehicles; (2) comprehensive feasibility of the hybrid energy storage system considering battery degradation reduction, energy storage price, operation cost, mass effect, placement, and future solid-state battery era; (3) parameter sensitivity of energy storage price and temperature; and (4) limitation discussion of this study.
	
	\subsection{Economy analysis}
	
\begin{figure}[!h]
		\centering
		\includegraphics[width=8cm]{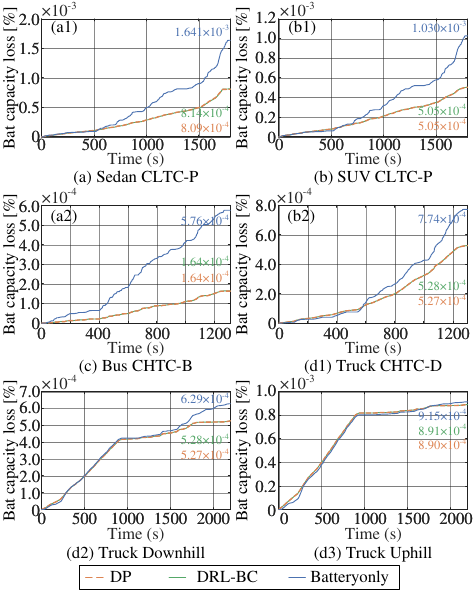}
		\caption{Comparative battery capacity loss results of battery-only, DRL-BC, and DP for (a) Sedan under CLTC-P, (b) SUV under CLTC-P, (c) Bus under CHTC-B, (d1) Truck under CHTC-D, (d2) Truck under Downhill cycle, (d3) Truck under Uphill cycle. Initial SC SoC equals 0.75.}
		\label{fig10_Qloss}
	\end{figure}

	To quantitatively evaluate the battery degradation reduction brought by the HESS, Fig. \ref{fig10_Qloss} presents the battery capacity loss results of different vehicles. The battery-only vehicle, HESS with the proposed DRL-BC, and HESS with offline discretized DP reference are compared.
	
	Across all six test scenarios, the battery degradation trajectories of the DRL-BC closely track the discretized DP references, indicating that the proposed online energy management strategy achieves near-optimal power allocation decisions in real time. 
	
	Under highly intermittent driving cycles, such as CLTC-P for Sedan/SUV in Fig. \ref{fig10_Qloss} (a/b), CHTC-B for the Bus in Fig. \ref{fig10_Qloss} (c), and CHTC-D for the Truck in Fig. \ref{fig10_Qloss} (d1), the battery capacity loss of DRL-BC is significantly reduced compared to the battery-only case. For instance, the battery capacity loss of the Bus is remarkably reduced from $5.76 \times 10^{-4}\%$ to $1.64 \times 10^{-4}\%$ (71.5\%), and the battery capacity loss reduction for the Sedan, SUV, and Truck under CHTC-D are 50.4\%, 51.0\%, and 31.8\%, respectively. This demonstrates that the supercapacitor pack effectively absorbs the high-frequency peak power associated with frequent acceleration and braking.
	
	Under continuous heavy-duty operations, the protective leverage of the HESS diminishes significantly. As depicted in Fig. \ref{fig10_Qloss} (d3), in the Truck Uphill cycle, the degradation trajectories of all three strategies nearly overlap, indicating that the limited energy capacity of the supercapacitor pack is rapidly depleted, forcing the battery pack to undertake the continuous high-power load regardless of the energy management strategy applied. For the Truck Downhill cycle in Fig. \ref{fig10_Qloss} (d2), the battery degradation can be reduced by 16.1\% with the DRL-BC strategy due to full-load energy feedback.
	
\begin{figure}[!h]
		\centering
		\includegraphics[width=8cm]{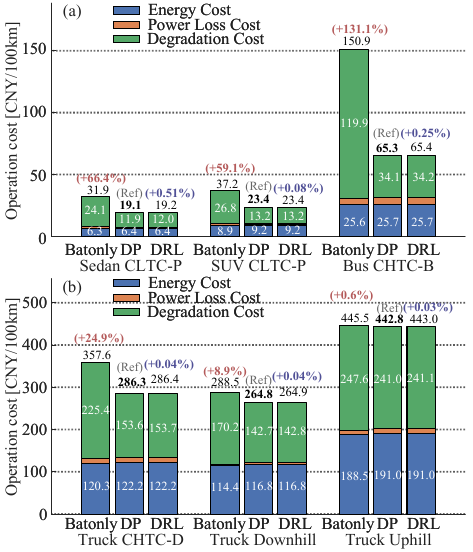}
		\caption{Economy analysis of battery-only, DRL-BC, and DP for (a) Sedan under CLTC-P, SUV under CLTC-P, Bus under CHTC-B and (b) Truck under CHTC-D, Truck under Downhill cycle, Truck under Uphill cycle. Initial SC SoC equals 0.75. Power loss cost can be calculated using subtraction.}
		\label{fig11_economy}
	\end{figure}

	Following the battery capacity loss evaluation, Fig. \ref{fig11_economy} further illustrates operational cost (CNY/100km) across different vehicle scenarios. The total operational cost is decomposed into traction energy cost, battery degradation cost, and power loss cost. The calculation is based on Eq. (\ref{eq:DP}) with a battery price of 756CNY/kWh and an electricity price of 0.6CNY/kWh. 
	
	The battery degradation cost dominates the total operation cost across all vehicle types, whether with battery-only or HESS. Meanwhile, the traction energy cost increase due to additional supercapacitor mass is also quantified. By introducing the HESS, the proposed DRL-BC strategy demonstrates substantial economic feasibility, particularly under driving cycles with high-frequency fluctuations, see Fig. \ref{fig11_economy} (a). 
	
	For the Bus under CHTC-B, the DRL-BC strategy significantly reduces the total operation cost from 150.9 to 65.4CNY/100km compared to the battery-only case, primarily driven by a reduction in degradation cost, from 119.9 to 34.2CNY/100km. For the Sedan and SUV, the total operation costs are 19.2 and 23.4CNY/100km with the DRL-BC strategy, tracking closely with the DP results. The near-optimal online performance is thereby demonstrated.
	
	However, for the Truck under the continuous heavy-duty Uphill cycle, the total operation cost decreases only slightly from 445.5 (battery-only) to 443.0CNY/100km (DRL-BC). The battery degradation cost remains high at 241.1CNY/100km, as the limited energy capacity of the supercapacitor pack is insufficient for sustained discharge. 
	
	Across the six scenarios, the operation cost obtained by DRL-BC is only 0.029–0.47\% higher than the discretized DP reference, demonstrating near-optimal online performance.
	
	While the operation cost comparisons validate the effectiveness of the energy management strategy, commercializing HESS requires a comprehensive techno-economic assessment. The initial cost of the supercapacitor and the severe spatial placement penalties must be weighed against these cost savings, which necessitate the multi-dimensional feasibility framework discussed below.

	\subsection{Feasibility analysis of the hybrid energy storage system}
	
	To present a comprehensive techno-economic assessment, a five-dimensional feasibility framework is established. Five crucial dimensions are mapped into a normalized feasibility score ranging from 0 to 100, where 100 consistently represents the most favorable engineering or economic condition. The mapping boundary conditions are defined as follows:
	
	\textbf{(1) Battery Degradation Reduction}: A positive metric normalized from $0\%$ (score 0) to $80\%$ (score 100), reflecting the percentage of battery degradation reduction by the DRL-BC energy management strategy.
	
	\textbf{(2) Operation Cost Reduction}: A positive metric normalized from $0\%$ (score 0) to $70\%$ (score 100), denoting the overall operation cost savings over the entire vehicle lifespan.
	
	\textbf{(3) SC Cost Efficiency}: An inverse metric mapping the SC/battery initial investment ratio, normalized from $40\%$ (score 0) to $0\%$ (score 100). A score approaching 100 indicates that the SC's initial investment is economically negligible compared to the battery pack.
	
	\begin{figure*}[!b]
		\centering
		\includegraphics[width=16cm]{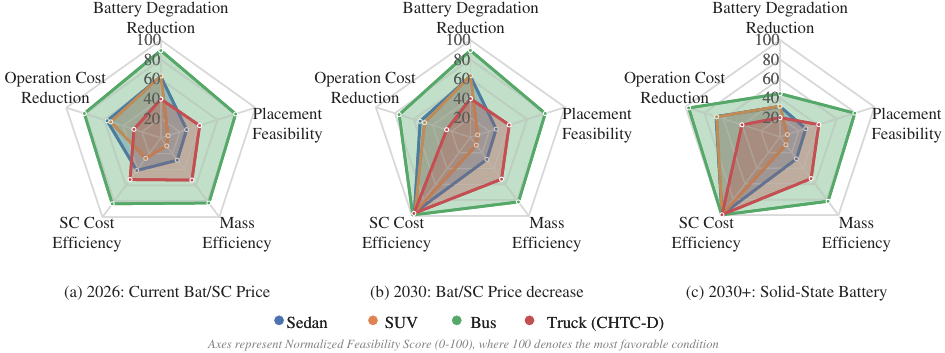}
		\caption{Feasibility analysis of HESS for Sedan, SUV, Bus, and Truck in (a) 2026, (b) 2030, and (c) 2030+ (Solid-state battery era). Five dimensions are compared: Battery Degradation Reduction, Operation Cost Reduction, SC Cost Efficiency, Mass Efficiency, and Placement Feasibility (HESS vs. battery only).}
		\label{fig12_Radar}
	\end{figure*}
	
	\textbf{(4) Mass Efficiency}: An inverse metric mapping the SC/battery mass ratio, ranging from $20\%$ (score 0) to $0\%$ (score 100). A score approaching 100 denotes negligible additional mass.
	
	\textbf{(5) Placement Feasibility}: An inverse metric mapping the SC/battery volume ratio, normalized from $45\%$ (score 0) to $0\%$ (score 100). A score approaching 0 indicates severe spatial placement constraints.
	
	The energy storage prices for batteries and supercapacitor from 2026 to 2030+ are listed in Table \ref{tab5} and are sourced from authoritative organizations. These values can influence the operation cost calculation.
	
	\begin{table}[!h] 
		\caption{Energy storage prices from 2026 to 2030+.}
		\label{tab5}
		\centering
		\begin{threeparttable}
			\resizebox{8cm}{!}{
				\begin{tabular}{llcc}
					\toprule
					Price of [CNY/kWh] & 2026 & 2030 & 2030+ (Solid-state battery era) \\
					\midrule
					Li-ion battery & 756$^1$ & 560$^2$ & -  \\
					Supercapacitor$^3$ & 28000 & 1400 & 1400  \\
					Solid-state battery$^4$ & - & - & 3500  \\
					\bottomrule
			\end{tabular}}
		\end{threeparttable}
		\begin{tablenotes}
			\footnotesize
			\item[1] $^1$ U.S. Department of Energy and BloombergNEF \cite{BatSCprice,Batprice1}.
			\item[2] $^2$ Goldman Sachs \cite{FutureBatPrice}. $^3$ EIT InnoEnergy \cite{SCprice1}. 
			\item[3] $^4$ Mordor Intelligence \cite{SSBprice}.
		\end{tablenotes}
	\end{table}
	
	Based on the above-defined criteria, Fig. \ref{fig12_Radar} illustrates the feasibility results of HESS across 2026, 2030, and 2030+ technological eras. For the Truck, only CHTC-D results are presented.
	
	Currently, in 2026, Fig. \ref{fig12_Radar} (a), the radar shapes are highly asymmetric. The Bus emerges as the only highly feasible candidate, exhibiting an excellent score across all five dimensions. This is primarily due to the low speed and frequent start/stop features of the driving cycle, as well as the negligible optimal SC size. The Sedan and SUV suffer from severe downgrades in both placement feasibility and SC cost efficiency. Despite theoretical battery life and operation cost savings, the significant volumetric penalty and high initial SC cost currently prevent HESS from being adopted in passenger vehicles. The Heavy-duty Truck (CHTC-D) maintains adequate placement feasibility and mass efficiency but is heavily constrained by moderate operation cost reduction and SC cost efficiency.
	
	\begin{figure*}[!t]
		\centering
		\includegraphics[width=16cm]{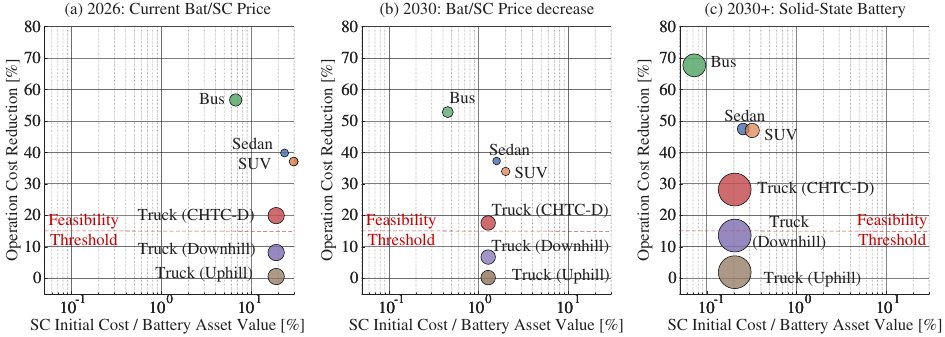}
		\caption{Techno-economic evolution and investment leverage shift of HESS across different vehicle types and scenarios in (a) 2026, (b) 2030, and (c) 2030+ (Solid-state battery era). The bubble size is proportional to the total monetary value of the protected battery pack. The red dashed line indicates a conservative 15\% commercial feasibility threshold to account for unmodeled hidden integration complexities and reliability risks.}
		\label{fig13_Bubble}
	\end{figure*}
	
	Moving to 2030, see Fig. \ref{fig12_Radar} (b). Due to a slight reduction in battery price, the operation cost reduction score also decreases slightly. The SC price collapse triggers an expansion of all polygons toward the SC cost-efficiency axis. However, these economic advantages expose the physical rigid bottlenecks: the mass efficiency and placement feasibility remain unchanged. The Sedan and SUV are still firmly constrained by their packaging; installing an SC pack at the chassis or in the trunk is still virtually unacceptable. The difference lies in the Truck (CHTC-D); even though the operation cost reduction is not significant, the SC cost efficiency makes the HESS feasible somehow.
	
	In the 2030+ solid-state battery era, we have the following assumptions: (1) the mass and volume energy density of the solid-state battery is twice that of the current LFP Li-ion battery; (2) the battery capacity loss model is the same as the current LFP Li-ion battery; (3) the battery C-rate is halved under the same power demand owing to the doubled battery capacity. Then, in Fig. \ref{fig12_Radar} (c), the feasibility results undergo an overall shift. Due to the extremely high cost of solid-state batteries, the relative SC investment approaches zero, resulting in SC cost efficiency approaching 100 across all vehicle types. However, under the assumed doubled energy density, the battery capacity is doubled and the corresponding C-rate is halved under the same power demand. Therefore, the relative battery degradation reduction provided by the HESS decreases.
	
	To further evaluate the economic leverage of the HESS, Fig. \ref{fig13_Bubble} illustrates the relationship between the relative initial SC/battery investment and the resulting operation cost reduction across different eras. The maximum SC cost efficiency in Fig. \ref{fig12_Radar} is less than 40\%. Considering the average vehicle service life of 12 to 15 years \cite{burnham2021comprehensive} and the simplified system dynamics, a $15\%$ operation cost reduction is set as a baseline feasibility threshold to account for real-world uncertainties. This 15\% value is a study-specific conservative benchmark rather than a universal industrial threshold; cases close to this boundary are therefore interpreted as marginally feasible rather than absolutely feasible or infeasible.
	
	In 2026, Fig. \ref{fig13_Bubble} (a), the initial cost of the supercapacitor constitutes a notable fraction of the battery asset value, ranging from approximately $7\%$ to nearly $30\%$ depending on the vehicle type. The Bus demonstrates the highest economic return, achieving an operation cost reduction of nearly $60\%$. Passenger vehicles (Sedan and SUV) also show operation cost reductions exceeding $30\%$, well above the 15\% threshold. However, heavy-duty applications under continuous load profiles, specifically Truck (Downhill) and Truck (Uphill), fall below the $15\%$ feasibility threshold, indicating that the economic return does not justify the initial investment under current pricing.
	
	As energy storage prices decrease in 2030, Fig. \ref{fig13_Bubble} (b), a distinct left shift is observed across all data points. The SC initial cost/battery asset value ratio drops to approximately $0.5\%$ to $2\%$, significantly reducing investment costs. While the relative positions of the vehicles on the y-axis remain stable, the operation cost reductions for Truck (Downhill) and Truck (Uphill) still remain below the $15\%$ threshold. This indicates that reducing the initial cost alone is insufficient to make HESS economically feasible for low-frequency, continuous-discharge scenarios.
	
	In the 2030+ solid-state battery era, shown in Fig. \ref{fig13_Bubble} (c), the bubbles enlarge, reflecting the increased economic value of the protected solid-state battery pack. Since the solid-state battery is expected to be considerably more expensive, the relative SC investment ratio falls below $1\%$, indicating a high economic return (with a small investment to protect a high-value asset). Then, the operation cost reduction for the Truck (CHTC-D) improves to over $20\%$, crossing the feasibility threshold entirely. Nevertheless, the Downhill and Uphill load scenarios remain marginal or infeasible. 
	
	Fig. \ref{fig13_Bubble} demonstrates that while future cost reductions and battery technology advancements will improve the investment leverage of HESS (shifting the data points to the left), the maximum achievable operation cost reduction remains strongly influenced by the load-frequency profile together with the vehicle/system configuration.
	
	\begin{figure}[!b]
		\centering
		\includegraphics[width=8cm]{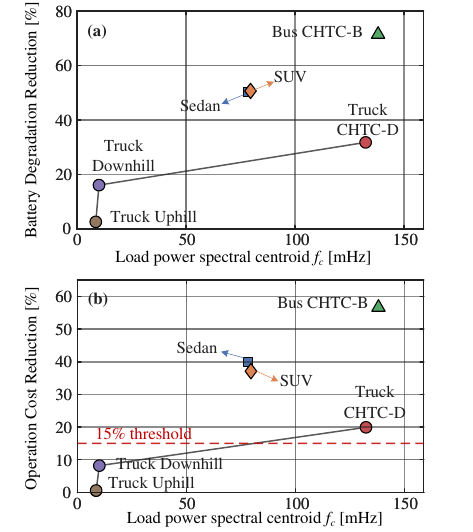}
		\caption{Quantified relationship between load-frequency characteristics and HESS benefits. The load-frequency characteristic is quantified by the load spectral centroid, $f_c$. (a) Relationship between $f_c$ and battery degradation reduction. (b) Relationship between $f_c$ and operation cost reduction. The connected Truck data points represent the same vehicle platform under the Uphill, Downhill, and CHTC-D driving cycles.}
		\label{fig14_4axis}
	\end{figure}

    To further quantify the influence of load characteristics, Fig. \ref{fig14_4axis} relates the load power spectral centroid, $f_c$, to the battery degradation reduction and operation cost reduction. For the same Truck platform, a clear monotonic trend can be observed: as $f_c$ increases from 8.6 mHz under the Uphill cycle to 10.0 mHz under the Downhill cycle and further to 132.4 mHz under CHTC-D, the battery degradation reduction increases from 2.6\% to 16.1\% and 31.8\%, while the operation cost reduction increases from 0.6\% to 8.2\% and 19.9\%, respectively. This indicates that, within the same vehicle platform, stronger load-frequency characteristics are associated with larger HESS benefits.
    
    However, the cross-vehicle comparison also shows that load-frequency characteristics are not the only determining factor. For example, although the Sedan and SUV have lower $f_c$ values than the Truck under CHTC-D, they still achieve higher battery degradation reduction and operation cost reduction. Therefore, the techno-economic benefit of the HESS is not governed by load-frequency characteristics alone, but jointly influenced by vehicle type, component sizing, and operating driving cycles.
	
	The study therefore concludes that HESS feasibility is jointly shaped by vehicle configuration and load-frequency characteristics. Within a given vehicle platform, higher-frequency transient loads generally favor HESS benefits, whereas sustained low-frequency loads tend to weaken its economic justification.
	
	It can be concluded that, for passenger vehicles, the HESS is expected in the future, i.e., the supercapacitor pack can be an optional premium configuration. For city buses, HESS is the most favorable application. For heavy-duty trucks, HESS deployment is not recommended in mining operations like mine shafts or mines, whereas it is highly feasible for urban and port transportation.
	
	\subsection{Parameter sensitivity analysis}
	
	\subsubsection{Influence of different temperatures}
	
	\begin{figure*}[!b]
		\centering
		\includegraphics[width=16cm]{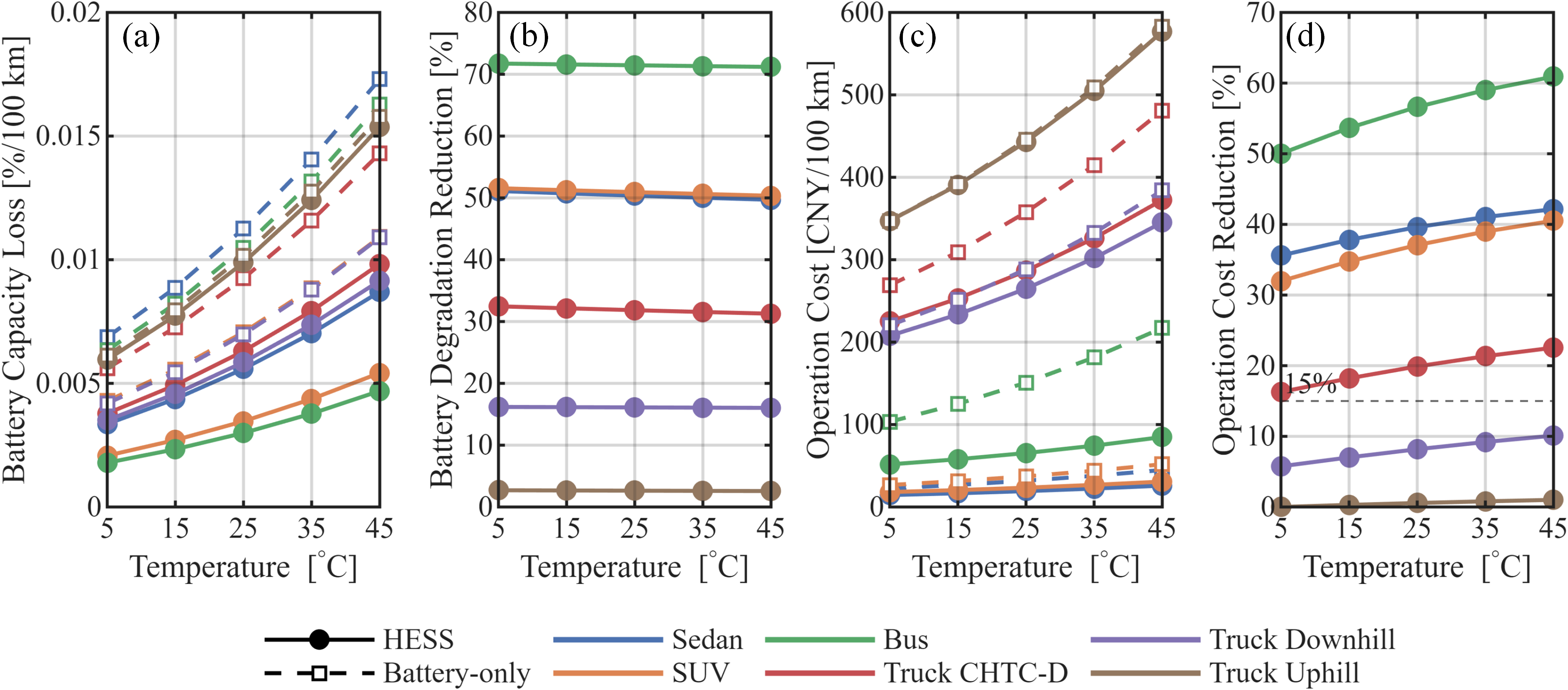}
		\caption{Battery temperature sensitivity analysis for different vehicles. (a) Battery capacity loss, (b) battery degradation reduction, (c) absolute operation cost, and (d) operation cost reduction under different temperatures.}
		\label{fig15_temperature_sensitivity}
	\end{figure*}
	
	\begin{figure*}[!t]
		\centering
		\includegraphics[width=16cm]{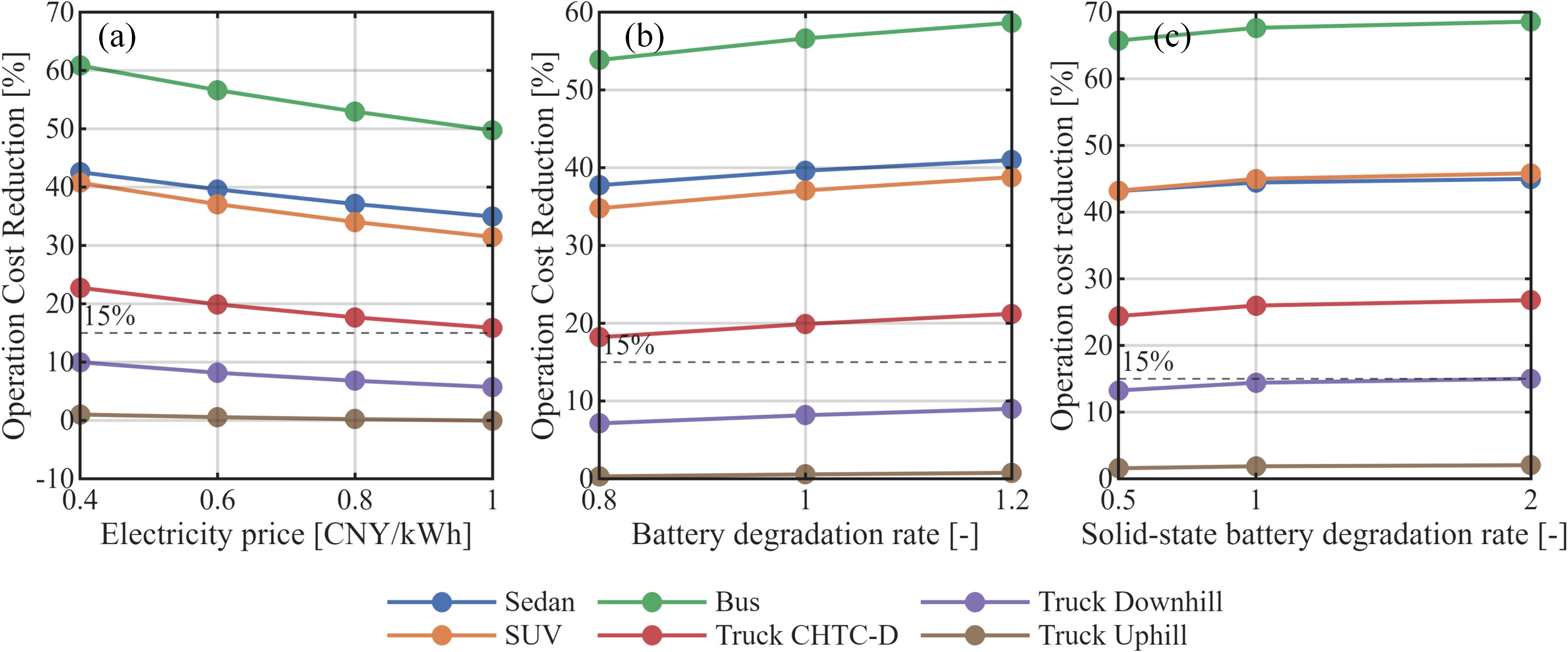}
		\caption{Sensitivity of operation cost reduction to (a) electricity price, (b) battery degradation rate uncertainty, and (c) solid-state battery degradation rate.}
		\label{fig16_other_sensitivity}
	\end{figure*}
	
	\begin{figure*}[!t]
		\centering
		\includegraphics[width=16cm]{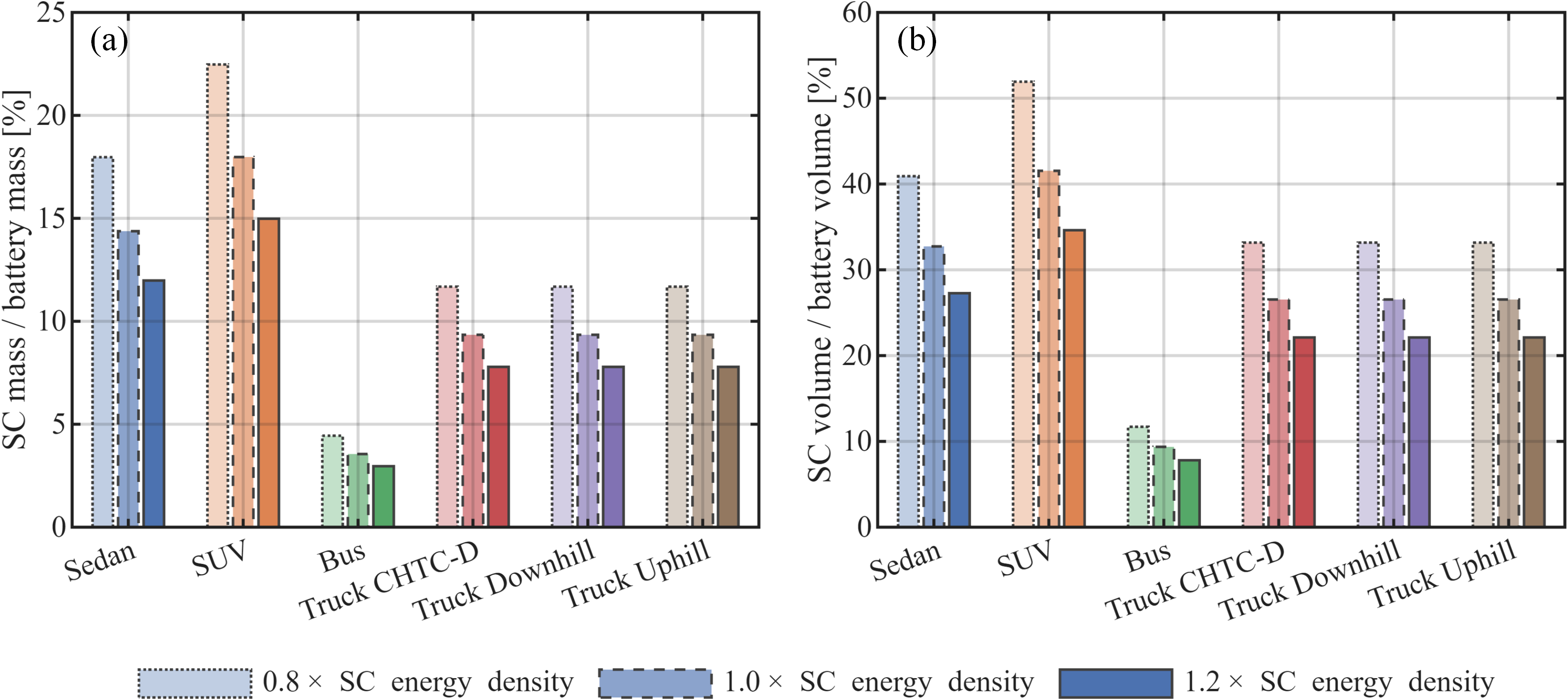}
		\caption{Sensitivity of the SC packaging penalty to SC energy density. (a) SC mass-to-battery mass ratio and (b) SC volume-to-battery volume ratio.}
		\label{fig17_SC_sensitivity}
	\end{figure*}
	
	Battery temperature $T_{bat}$ in Eq. (\ref{eq:Qloss}) directly affects the battery degradation rate and may therefore influence the economic benefit attributed to the HESS. As shown in Fig. \ref{fig15_temperature_sensitivity} (a, b), the absolute battery capacity loss [\%/100km] of both the battery-only and HESS cases increases substantially as the temperature increases from 5 to 45~$^{\circ}$C. Nevertheless, the relative benefit of the HESS is considerably less sensitive to temperature. The battery degradation reduction remains approximately 50--52\% for the Sedan and SUV, around 71--72\% for the Bus, and around 31--32\% for the Truck under CHTC-D throughout the investigated temperature range. The corresponding reductions for the Truck under Downhill and Uphill driving cycles remain around 16\% and 3\%, respectively. This indicates that temperature strongly changes the absolute battery capacity loss but has limited influence on the relative battery protection capability of the HESS solution.
	
	In Fig. \ref{fig15_temperature_sensitivity} (c, d), the economic benefit becomes more significant at higher temperatures since battery degradation accounts for a larger proportion of the operation cost. Accordingly, the operation cost reduction increases with temperature for all investigated cases, while the relative ranking among the vehicle applications remains essentially unchanged. In particular, the Bus consistently exhibits the largest economic benefit, whereas the Truck Uphill and Downhill cases remain economically unfavorable. These results suggest that the principal cross-vehicle feasibility conclusions are robust to the considered battery temperature variation.
	
	\subsubsection{Influence of economic and degradation parameters}
	
	Fig. \ref{fig16_other_sensitivity} (a), (b), and (c) further examine uncertainties associated with electricity price, battery degradation rate, and solid-state battery performance, respectively. Increasing the electricity price from 0.4 to 1.0~CNY/kWh reduces the relative operation cost benefit of the HESS since the traction energy and power loss costs become more important compared with the battery degradation cost saving. Nevertheless, the main feasibility ranking remains unchanged: the Bus maintains the highest operation cost reduction (over 50\%), followed by the Sedan and SUV (over 30\%). Similarly, varying the battery degradation rate from 0.8 to 1.2 times the baseline produces only moderate changes in operation cost reduction and does not alter the overall conclusions, see Fig. \ref{fig16_other_sensitivity} (b). The corresponding battery degradation reduction changes only slightly because the same degradation rate uncertainty is applied consistently to both the HESS and battery-only cases.
	
	The degradation behavior of future solid-state batteries remains subject to considerable technological uncertainty, given the diversity and continuing development of SSB chemistries and designs \cite{joshi2025comprehensive}. Therefore, retaining the doubled battery capacity and the resulting halved C-rate, the solid-state battery degradation kinetics are varied from 0.5 to 2 times the baseline value. As shown in Fig. \ref{fig16_other_sensitivity} (c), the economic advantage of HESS increases moderately as the degradation rate increases since the value of battery protection becomes greater. However, the relative cross-vehicle trend is preserved over the considered range: the Bus remains the most favorable application, while the Truck Downhill case approaches the 15\% feasibility threshold only under the highest assumed degradation rate, and the Truck Uphill case remains economically unfavorable. Thus, uncertainties in the considered economic and degradation parameters affect the magnitude of the estimated benefits but do not fundamentally change the overall cross-vehicle conclusions.
	
	\subsubsection{Influence of supercapacitor energy density}
	
	The robustness of the physical sizing conclusions is further examined by varying the mass and volume energy density of the SC from 0.8 to 1.2 times the baseline value while maintaining the same electrical sizing and EMS results. As shown in Fig. \ref{fig17_SC_sensitivity}, even under the conservative 0.8-times energy-density case, the maximum SC-to-battery mass ratio is approximately 22\% for the SUV, remaining below the 30\% engineering screening boundary. The corresponding maximum volume ratio is approximately 52\%, also substantially below the 100\% volume boundary. The Bus consistently exhibits the smallest mass and volume penalties, whereas the SUV remains the most packaging constrained application. The three Truck driving conditions share the same optimized hardware configuration and therefore have identical physical sizing sensitivity. Overall, reasonable variations in SC energy density change the magnitude of the packaging penalty but do not alter the principal feasibility conclusions.
	
	\subsection{Limitation and discussion}
	
	This study is intended as a cross-vehicle techno-economic feasibility framework rather than a production HESS design. Several model simplifications should therefore be acknowledged: The longitudinal vehicle model neglects rotating component inertia, axle load transfer, and tire--road slip. The battery capacity loss model assumes a spatially uniform pack temperature and an approximately uniform current distribution among parallel-connected cells. Calendar aging, depth-of-discharge influence, cell-to-cell inconsistency, low-temperature degradation mechanisms such as lithium plating, and SC self-discharge are not explicitly modeled. The adopted capacity loss model was developed and calibrated using earlier-generation LFP cells \cite{song2015optimization} and may not quantitatively reproduce the degradation behavior of modern blade-type LFP batteries. It is therefore used here primarily for consistent cross-vehicle comparative evaluation rather than cell-specific lifetime prediction. Also, the SC parameters are representative of the electrical double-layer capacitor technologies. The applicability to other battery chemistries, operating SoC windows, and supercapacitor technologies is consequently subject to uncertainty. The sensitivity analyses presented above are intended to quantify part of these uncertainties rather than eliminate them.
	
	In addition, the mass and volume limits adopted in the sizing procedure should be interpreted as engineering feasibility boundaries rather than universal industrial standards. Although the major DC/DC converter mass and envelope volume are estimated and discussed, the converter purchase cost and auxiliary components such as external filters, contactors, cooling manifolds, busbars, protection devices, and installation clearances are not modeled in detail; therefore, the perspective views in Fig. \ref{fig4_perspective_view} represent conceptual packaging solutions rather than production designs. The 2030+ solid-state battery analysis is likewise a scenario-based projection based on assumed energy density, cost, and degradation behavior, rather than a prediction of a specific future chemistry. Finally, although the same-vehicle multi-cycle comparisons help reveal the influence of load-frequency characteristics, vehicle configuration, component sizing, and operating conditions may still have coupled effects. Hardware experiments and vehicle-level demonstrations are therefore required before the proposed feasibility boundaries can be translated into specific industrial designs.
	
	\section{Conclusion \label{conclusion}}

	This study establishes a multi-dimensional techno-economic evaluation framework to systematically assess the feasibility of HESS in electric vehicles. By integrating cross-vehicle optimal sizing, mass-volume constraints, online energy management, and life-cycle feasibility matrices, the main conclusions are drawn as follows:
	
	The life-cycle economic benefits of supercapacitors are highly sensitive to physical mass-volume penalties. Currently, city buses are the most promising vehicle type for HESS applications, capable of mitigating battery degradation at very low additional cost. Conversely, standard passenger vehicles and heavy-duty trucks for urban/port scenarios remain constrained by spatial or heavy-load penalties; their HESS adoption will become feasible given a significant future reduction in supercapacitor prices.
	
	The quantified analysis further shows that, within the same vehicle platform, stronger load-frequency characteristics are associated with larger HESS benefits. Across different vehicle platforms, however, the final techno-economic outcome is jointly determined by load-frequency characteristics and vehicle/system configuration. Vehicles with highly transient, high-frequency power demands generally exhibit larger HESS benefits, whereas relatively continuous load profiles tend to reduce the economic feasibility of HESS.
	
	The economic benefits of HESS will diminish as the price collapse and technological evolution of batteries unfold. Looking toward the 2030+ era, extending battery lifespan and energy density will further compress the need for onboard supercapacitors. Nevertheless, the high initial costs of solid-state batteries significantly amplify the asset protection leverage of degradation mitigation. By protecting the high-value batteries, the HESS remains an interesting solution for future electric vehicles.
	
	Note that the vehicle dynamics simplifications may affect the absolute traction power estimation and should therefore be considered when translating the identified feasibility boundaries to production vehicles. Several future directions are still interesting: (1) Future studies could integrate thermal safety dynamics into the economic evaluation. Supercapacitors inherently reduce peak currents, which contributes to reducing battery heat generation. This capability not only relieves the operational load on the vehicle thermal management system but also offers specific functional advantages in sub-zero climates, where supercapacitors can mitigate cold-weather battery constraints, e.g., lithium plating and power fade. (2) Guided by the load-frequency dependence, future research could explore the applicability of HESS in extreme-duty and off-road sectors. Vehicles such as electric racing vehicles, heavy construction machinery, or tracked vehicles are characterized by highly transient, stochastic power demands resulting from rapid accelerations, variable operational resistances, or skid-steering mechanisms. In these specialized applications, the frequent load fluctuations can amplify the asset protection leverage of supercapacitors, presenting a practical scenario for future HESS integration.
	
	\section*{Supplementary Materials}
	
	Additional information supporting this study is provided in the Supplementary Materials. The supplementary file includes detailed parameters of motor efficiency maps, implementation details of the DRL-BC energy management strategy, training settings, convergence analyses, ablation results, rule- and MPC- based energy management strategies, and additional results for energy management comparisons.
	
%
	
%
	
	\section*{Acknowledgment}
	
	This research was supported in part by the National Natural Science Foundation of China (Grant No. 52507161) and Natural Science Foundation of Hunan Province (Grant No. 2025JJ60823).
	
	
	
	

	\bibliographystyle{elsarticle-num}
	\bibliography{Reference}
	
\end{document}